\documentstyle[11pt,aaspp4]{article}

\begin{document}
\title{Low-Mass Star Formation and the Initial Mass Function in the
$\rho$~Ophiuchi Cloud Core\altaffilmark{1}}
\author{K. L. Luhman\altaffilmark{2} \& G. H. Rieke}

\affil{kluhman@as.arizona.edu, grieke@as.arizona.edu}
\affil{Steward Observatory, The University of Arizona, Tucson, AZ 85721}

\altaffiltext{1}{Observations reported in this paper were obtained with
the Multiple Mirror Telescope operated by the Smithsonian Astrophysical
Observatory and the University of Arizona.}

\altaffiltext{2}{Current address: Harvard-Smithsonian Center for Astrophysics, 
60 Garden St., Cambridge, MA 02138.}

\begin{abstract}

We have obtained moderate-resolution ($R=800$-1200) $K$-band spectra for
$\sim100$ stars within and surrounding the cloud core of $\rho$~Oph. 
We have measured spectral types and continuum veilings and have combined 
this information with results from new deep imaging. Using the latest 
evolutionary tracks of D'Antona \& Mazzitelli to interpret the H-R diagram
for $\rho$~Oph, we infer ages ranging between 0.1 and 1~Myr for the Class~II 
and III sources (i.e., those that have emerged from their natal cocoons). 
A few stars may be slightly older. The IMF peaks at about 0.4~$M_\odot$ 
and slowly declines to the hydrogen burning limit with a slope of $\sim-0.5$ 
in logarithmic units (Salpeter is $+1.35$). Our lower limits on the numbers of 
substellar objects 
demonstrate that the IMF probably does not fall more steeply below the hydrogen 
burning limit, at least down to $\sim0.02~M_\odot$. The derived IMF is 
consistent with previous findings that the $\rho$~Oph IMF is roughly flat 
from 0.05 to 1~$M_\odot$. The exact shape of the mass function remains
a function of the theoretical evolutionary tracks and, at the lowest masses,
the conversion from spectral types to effective temperatures.

We then make the first comparison of mass functions of stars and pre-stellar
clumps (Motte, Andr\'{e}, \& Neri) measured in the same region.  The similar
behavior of the two mass functions in $\rho$~Oph supports the suggestion of
Motte et al.\ and Testi \& Sargent that the stellar mass function 
in young clusters is a direct product of the process of cloud fragmentation.

We have also studied the very young and often still embedded Class~I and 
flat spectrum objects.  After considering the effect of extinction on the 
SED classifications of the sample, we find that $\sim17$\% of the 
$\rho$~Oph stars are Class~I, implying $\sim0.1$~Myr for the lifetime of 
this stage.  In spectra separated by two years, we observe simultaneous
variability in the Br$\gamma$ emission and $K$-band continuum veiling
for two stars, where the hydrogen emission is brighter in the more heavily 
veiled data. This behavior indicates that the disk may contribute significantly
to continuous $K$-band emission, in contrast to the proposal that the infalling 
envelope always dominates. 
Our detection of strong 2~\micron\ veiling ($r_K=1$-4) in 
several Class~II and III stars, which should have disks but little envelope 
material, further supports this proposition.
We also detect absorption features in the spectra of $\sim25$\% of 
Class~I and flat-spectrum sources, demonstrating the feasibility of studying 
the photospheres of extremely young protostars. 

\end{abstract}


\section{Introduction}

The birth of low-mass stars and brown dwarfs as well as the evolution of 
circumstellar disks and planetary systems can best be studied in the nearest 
($d<500$~pc) young (age $<10$~Myr) stellar clusters and associations. For
example, the distribution of masses in these populations directly reflects 
the initial mass function (IMF). In contrast, in regions like the local 
field or globular clusters, the present day mass function is the product 
of a complex evolution. 

Until recently, these advantages of young clusters were difficult to 
exploit because of extinction. However, by complementing optical techniques 
with modern infrared (IR) imaging and spectroscopy, we can estimate 
luminosities and spectral types for complete, well-defined samples of 
young objects within obscured star forming regions. Due to the luminous 
nature of newborn substellar objects, such data can be converted
to mass functions that reach below the hydrogen burning limit.  

The $\rho$~Ophiuchi dark cloud core contains a very young ($<1$~Myr),
nearby ($d=160$~pc), compact ($D\sim20\arcmin$) population of $\sim100$
low-mass stars, making it ideal for IMF studies. Due to the large extinction 
within the cloud, optical spectral 
types and photometry (Bouvier \& Appenzeller 1992; Mart{\'\i}n et al.\ 1998)
and soft X-ray data (Montmerle et al.\ 1983; Casanova et al.\ 1995)
are available for only a small fraction of the stars.  Until recently, the 
$\rho$~Oph members were studied primarily through their near- to mid-IR 
spectral energy distributions (SEDs), which were observed with single element 
detectors on the ground and aboard the {\it IRAS} satellite
(Elias 1978; Lada \& Wilking 1984; Wilking, Lada, \& Young 1989; Greene et 
al.\ 1994).  Following the initial observations of $\rho$~Oph 
with near-IR arrays (Barsony et al.\ 1989; Rieke, Ashok, \& Boyle 1989), Greene 
\& Young (1992) surveyed 650~arcmin$^{2}$ of the cloud at $J$, $H$, and $K$, 
providing a relatively sensitive ($K\sim13$) photometric census of the stellar 
population.  To reach young objects below the hydrogen burning limit, 
Rieke \& Rieke (1990) obtained deep images ($K\sim15.5$) of 200~arcmin$^{2}$ 
within the cloud core ($A_{V}\gtrsim50$, Wilking \& Lada 1983)
and identified several brown dwarf candidates. The substellar nature of
these sources has been supported by {\it ISO} mid-IR photometry (Comer\'{o}n et
al.\ 1998, hereafter CRCTL) and confirmed through spectroscopy in the optical 
(Luhman, Liebert, \& Rieke 1997) and IR (Williams et al.\ 1995; 
Wilking, Greene, \& Meyer 1998, hereafter WGM).  Comer\'{o}n et al.\ (1993) 
(hereafter CRBR) supplemented the data of Rieke \& Rieke (1990) with 
$JHKL\arcmin$ aperture photometry of a complete sample
of sources projected against the dense cloud core.  They introduced a method
of fitting the data to estimate individual source properties and deduced 
a logarithmically flat IMF (the Salpeter index is $+1.35$) down to and below 
the hydrogen burning limit. Strom, Kepner, \& Strom (1995) used an alternate 
approach introduced by Meyer (1995) that also indicated a flat IMF. 
Williams et al.\ (1995) showed that the techniques of CRBR and Strom et 
al.\ (1995) agree at the detailed level of individual source properties as 
well as overall IMF slope. 

Following these photometric studies, near-IR spectroscopy has been applied 
in studying other properties of the embedded stellar population in $\rho$~Oph.
After measuring $K$-band spectral types for 19 stars ($K<10.5$, $R=500$),
Greene \& Meyer (1995) constructed a Hertzsprung-Russell (H-R) diagram
in which the theoretical evolutionary tracks of D'Antona \& Mazzitelli (1994) 
(hereafter DM94) implied an age of $\sim0.5$~Myr for the cluster.
Greene \& Lada (1996) also observed $\sim100$ young stars in Taurus and 
$\rho$~Oph ($R=500$) with broader wavelength
coverage (1.15-2.42~\micron).  The appearance of the spectra was
closely correlated with SED class, indicating, as expected, that reddening 
increases and photospheric features weaken 
from Class~III to Class~I sources (Lada 1987) and thus confirming the previous 
conclusions drawn from spectroscopy of 2.3~\micron\ CO absorption in young 
$\rho$~Oph members (Casali \& Matthews 1992). This trend is consistent
with the large obscuration and strong continuum veiling expected from
circumstellar material in the presumably younger, earlier SED classes.
Finally, Casali \& Eiroa (1996) and Greene \& Lada (1997) used high 
resolution spectra ($R=17,000$ and 21,000, respectively)
to measure continuum veilings and line widths for a small sample
of flat-spectrum sources in $\rho$ Oph and other young clusters. They 
demonstrated that the line profiles are consistent with absorption originating 
in very young stellar photospheres rather than circumstellar disks, an
important result that supports the validity of our classification of 
young stellar populations through IR spectra.

In this paper, we expand on the previous IR spectral classification in 
$\rho$~Oph and continue a program started in L1495E
(Luhman \& Rieke 1998, hereafter LR) and IC~348 (Luhman et al.\ 1998b,
hereafter LRLL).  We have obtained $K$-band spectroscopy of $\sim100$ 
sources, combining a magnitude-limited sample in the cloud core ($K\lesssim12$) 
with a representative population from the outer regions of the cluster
($K\lesssim11$). The
moderate spectral resolution ($R=800$-1200) of the data allows both reasonably 
accurate spectral types ($\pm1$-2 subclasses) and sensitivity to low-mass 
members of $\rho$~Oph ($\geq0.1$~$M_{\odot}$). With the measured spectral types 
we construct an H-R diagram and use theoretical evolutionary tracks to derive 
an IMF and star formation history for the emerged members of the cluster
(predominantly Class~II and III). Likely substellar cloud 
members examined by Luhman et al.\ (1997), WGM, and CRCTL are added to the 
mass function, extending the IMF to lower masses. We then use a new $K$-band 
luminosity function (KLF) of the cloud core to develop a model of the reddened 
background population and identify likely cluster members lacking spectroscopy. 
After accounting for these sources, we arrive at an IMF complete to
$\sim0.08$~$M_\odot$, which is compared to results from other young clusters. 

Using the mid-IR photometry of previous authors, we revise the SED 
classifications of the $\rho$~Oph population by dereddening the near- to 
mid -IR spectral indices with guidance from our near-IR spectra.
With the new classifications, the relative numbers of sources 
in the different categories are consistent with expectations that 
the dust embedded stage should have a very short duration. 
The broad-band photometric data combined with the measurements of
the $K$-band continuum veiling and Br$\gamma$ emission present
new constraints on the nature of the IR excess emission in young stars.  
In particular, simultaneous variability of Br$\gamma$ emission and veiling
is seen in two stars, possibly indicating the importance of disks as 
sources of 2~\micron\ excess emission. In addition, we find that 
photospheric absorption features are detectable in a number of Class~I and 
flat-spectrum sources, while significant veiling is measured in several
Class~II objects.  Both the placements on the H-R diagram and the distributions 
of SED classes indicate a slight increase in average age between the 
cloud core and surrounding region. 

\section{Observations}
\label{sec:oph.obs}

\begin{figure}
\plotfiddle{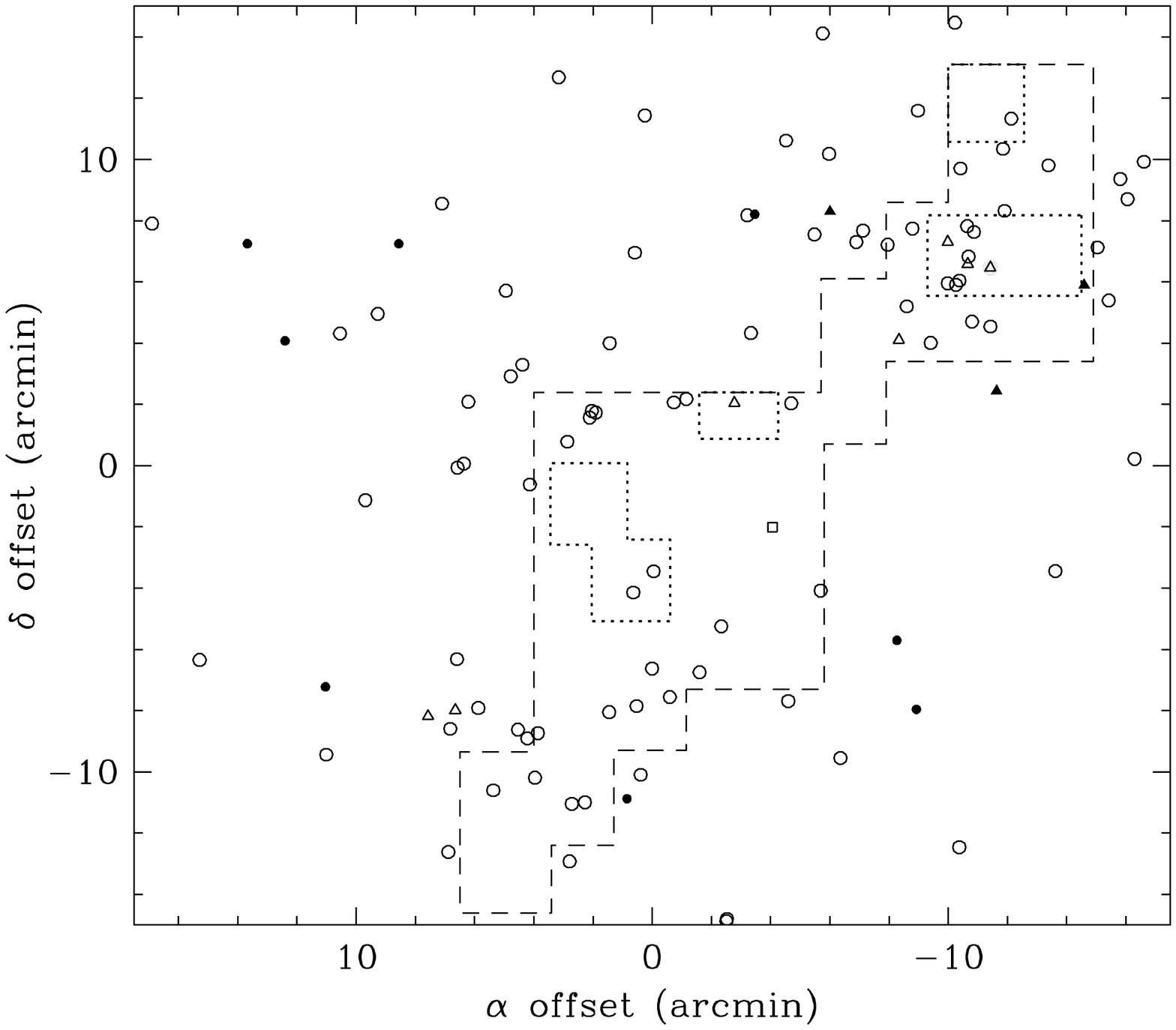}{6.5in}{0}{80}{70}{-245}{0}
\caption[Finding chart for the spectroscopic sample in $\rho$~Oph]{
All sources in $\rho$~Oph observed with $K$-band spectroscopy, with the 
exception of seven objects falling outside of this region 
(IRS2, IRAS64A, SR22, WSB38, WSB45, WSB46, ROXs39). Filled circles represent
stars classified as background giants in this work and open circles are all
other sources in our sample. Open triangles and the open square are
the additional low-mass objects observed spectroscopically by WGM and 
Luhman et al.\ (1997). Filled triangles are identified as likely
background stars by WGM.  The area within the dashed line was surveyed through
IR images by CRBR and corresponds to the cloud core ($A_{V}\gtrsim50$; Wilking
\& Lada 1983).  Deeper $K$-band
photometry of the same region was obtained for this work, in addition to 
$J$ and $H$ images towards the positions outlined by dotted lines. 
The origin corresponds to $\alpha=16^{\rm h}24^{\rm m}07\fs7$,
$\delta=-24\arcdeg24\arcmin00\arcsec$ (1950).
} 
\label{fig:oph.map}
\end{figure}

We performed $K$-band spectroscopy on sources in $\rho$~Oph using the 
near-IR long-slit spectrometer FSpec (Williams et al.\ 1993) at the Steward 
2.3~m Bok Reflector on Kitt Peak on 1994 July 15 and 1995 May 16, 17, 19
and at the Multiple Mirror Telescope (MMT) on Mount Hopkins on 1994 July 2-3,
1996 May 28-31, and 1996 June 3-6.  Most of the spectra were obtained
in 1996 with a grating providing a two-pixel resolution of 
$R=\lambda/\Delta\lambda=1200$, while the data from 1994 and 1995 have a
resolution of $R=800$.  The observations and data reduction 
procedures were identical to those described by LRLL. 
We selected for spectroscopy $\sim30$ sources with $K\lesssim12$ appearing 
in IR images of the cloud core by CRBR. With the addition of spectra from
Williams et al.\ (1995), our sample contains $\sim40$ sources and is 
nearly complete to $K=12$. In the 
region surrounding the core and within Figure~\ref{fig:oph.map}, we 
observed $\sim70$ sources with $K\lesssim11$ found in Greene \& Young (1992).
A few bright stars outside of the region in Figure~\ref{fig:oph.map} were also 
observed, taken from Wilking, Lada, \& Young (1989) and Greene et al.\ (1994). 
 
\begin{figure}
\plotfiddle{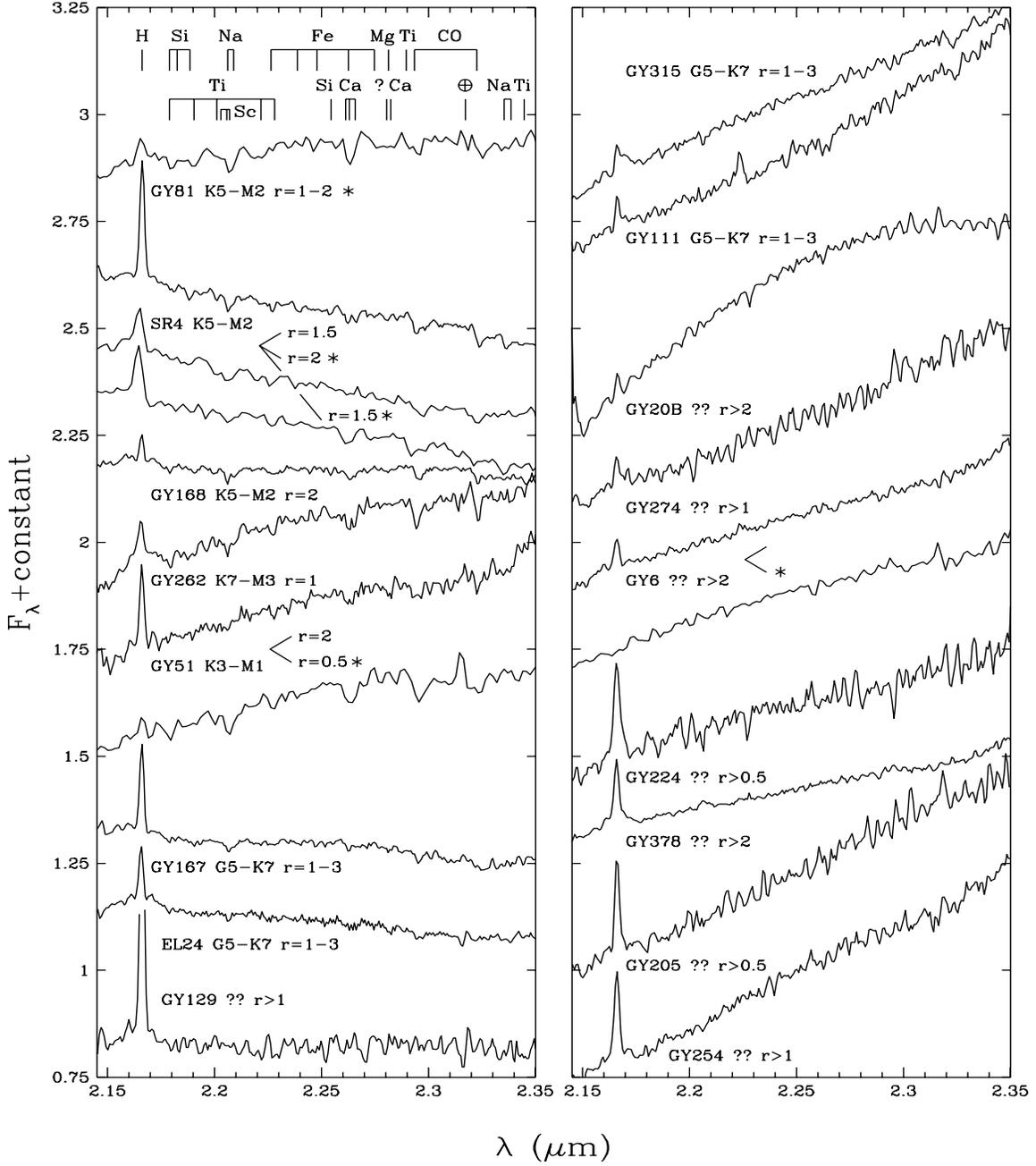}{6.5in}{0}{80}{70}{-245}{0}
\caption[$K$-band spectra of sources in $\rho$~Oph ($r_{K}\geq1$)]{
$K$-band spectra at $R=1200$ of heavily veiled sources ($r_{K}\geq1$) 
with additional data at $R=800$ ($\ast$).
The curvature in the spectrum of GY20B is not real and is due to difficulties
in extracting the spectrum separately from the companion GY20A.
Spectra are normalized at 2.2~\micron\ with constant offsets.
}
\label{fig:oph.irspec4}
\end{figure}

The spatial distribution of the spectroscopic sample is illustrated in 
Figure~\ref{fig:oph.map} and the spectra are presented in 
Figs.~\ref{fig:oph.irspec1}-\ref{fig:oph.irspec5}.  Line identifications were 
made by comparison to the high-resolution ($R\geq45,000$) spectra of cool stars
obtained by Wallace \& Hinkle (1996). 

Imaging at $K_s$ (1.9-2.3~\micron) 
was performed at the Bok Reflector on the nights of 1998 
April 12-15 using the Steward Observatory NICMOS3 $256\times256$ near-IR 
camera at a plate scale of 0\farcs64~pixel$^{-1}$, corresponding to 
a total field of $2\farcm7\times2\farcm7$.  In 43 pointings, we observed the
same region covered by CRBR (outlined by the dashed lines
in Figure~\ref{fig:oph.map}), in addition to a
few extra positions, for a total of $\sim250$~arcmin$^{2}$. Followup 
observations were made with standard $J$ and $H$ during 1998 May~8-10 
and July~14.  We selected 
pointings that contained several faint, possibly low-luminosity 
$K$-band sources (outlined by dotted lines in Figure~\ref{fig:oph.map}).
At each position, we obtained images 
with dithers of 5\arcsec\ in a $4\times4$ grid, facilitating
efficient flat-fielding and removal of bad pixels. Each exposure 
consisted of two coadded 30~s frames, producing a total exposure
time of 16~minutes. A flat field was constructed by median combining 
frames that lacked bright sources.  For calibration, we observed
a photometric standard star, SR3, periodically during
the night.  These exposures were taken out of focus to guard against 
saturation. After dark-subtracting and flat-fielding the images, the 16 frames
at a given position were shifted and combined into one image.  The tasks
DAOFIND and PHOT under the package APPHOT were used in measuring the stellar
coordinates and extracting photometry.  The density of stars was low enough
that we could measure the background emission in an annulus around each star
and subtract it from the photometry. The data were calibrated with the 
CIT photometry of SR3 from Elias et al.\ (1982) and are essentially in the 
CIT system. The $K$ images are complete to $K\sim17$, based both on the 
detection of many sources down to $K=17.5$ and, more importantly, 
because the background source counts do not deviate from the expected 
behavior until $K>17$ (see \S~\ref{sec:oph.complete}). 
Using the coordinates measured by Barsony et al.\ (1997, hereafter BKLT) for
stars above their detection limit ($K<14.5$) and below our saturation
limit ($K>10$), we derived relatively precise 
($\pm1\arcsec$) coordinates for the fainter sources ($K>14.5$).  

\section{Individual Source Characteristics}

\subsection{Spectral Types}
\label{sec:oph.irclass}

The spectral types derived for the $\rho$~Oph sample are listed in Table~1, 
along with measurements of Br$\gamma$ emission and continuum veiling at 
2.2~\micron\ ($r_{K}=I_{2.2}({\rm IR~excess})/I_{2.2}({\rm star}))$.
Typical uncertainties are 0.25-0.5 in $r_{K}$, depending on the signal-to-noise,
spectral type, and degree of veiling. Most of the objects in our sample
have been classified by the metallic absorption lines appearing in the $K$-band
spectra, as described in \S~\ref{sec:oph.kband}.
In our $K$-band sample, we include all additional data obtained by
Williams et al.\ (1995) (GY5, GY10, GY37, GY202, 162618-242611). 
A total of 40 sources from this combined spectroscopic sample fall within the
cloud core.
We have adopted the spectral type of WGM for one of these (162618-242416)
and now refer to it as part of the WGM sample.
In \S~\ref{sec:oph.steam}, we derive spectral types from the steam absorption in
these spectra and compare our results to those of 
WGM. We revise the classifications of WGM for objects not in our sample to 
correct for a possible systematic error due to gravity dependence 
of the steam band strength. Comments on the individual sources are given in 
\S~\ref{sec:oph.irappend}. The optical classification of the 
brown dwarf GY141 (Luhman et al.\ 1997) is also included in Table~1.

\subsubsection{IR Spectral Classification of the New Sample}
\label{sec:oph.kband}

\begin{figure}
\plotfiddle{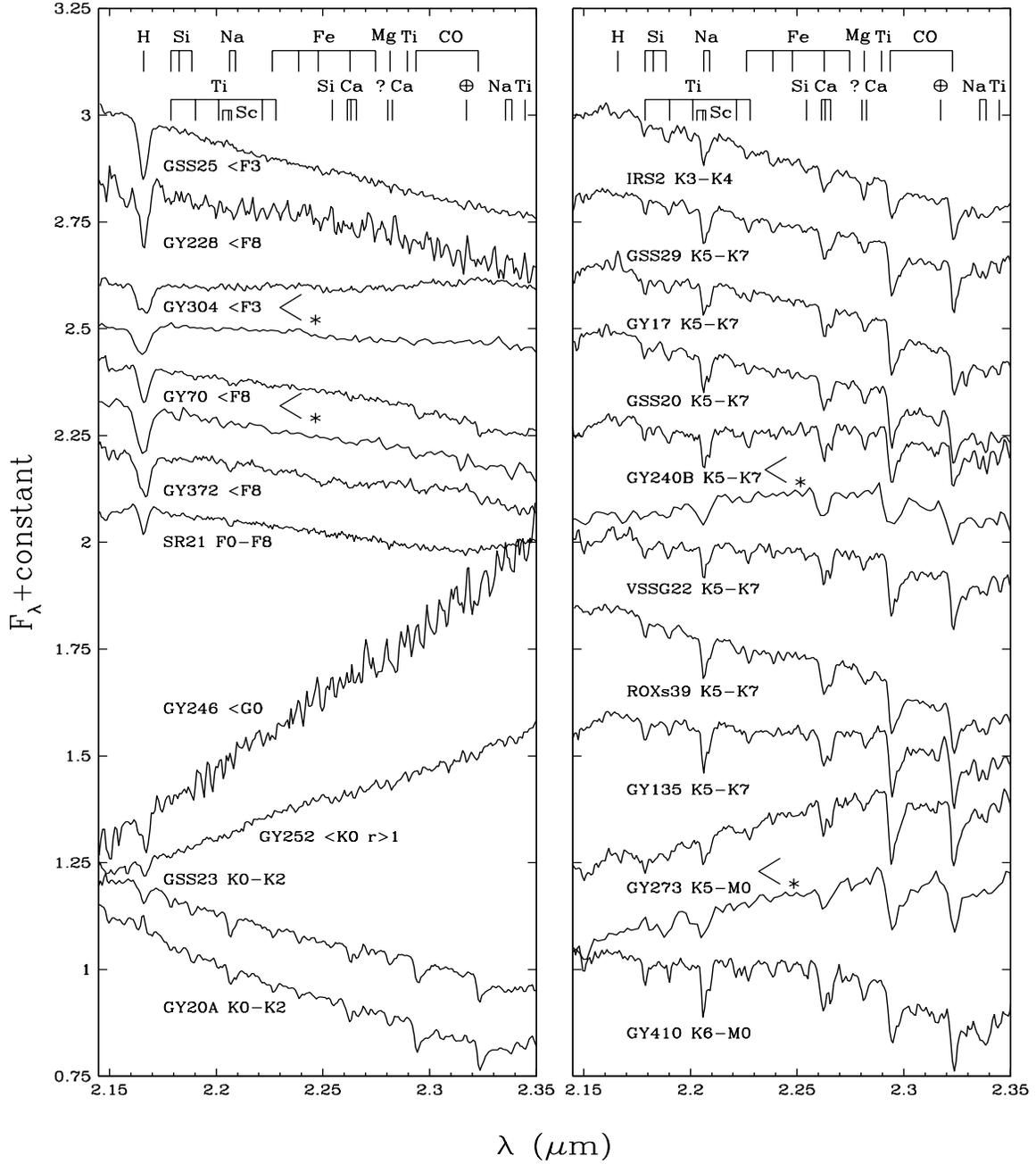}{6in}{0}{80}{70}{-245}{-20}
\caption[$K$-band spectra of sources in $\rho$~Oph ($<$F8 or $r_{K}<0.25$)]{
$K$-band spectra at $R=1200$ of early-type stars and sources exhibiting
no detectable continuum veiling ($r_{K}<0.25$), with additional data at
$R=800$ ($\ast$). Spectra are normalized at 2.2~\micron\ with constant offsets.
}
\label{fig:oph.irspec1}
\end{figure}

\begin{figure}
\plotfiddle{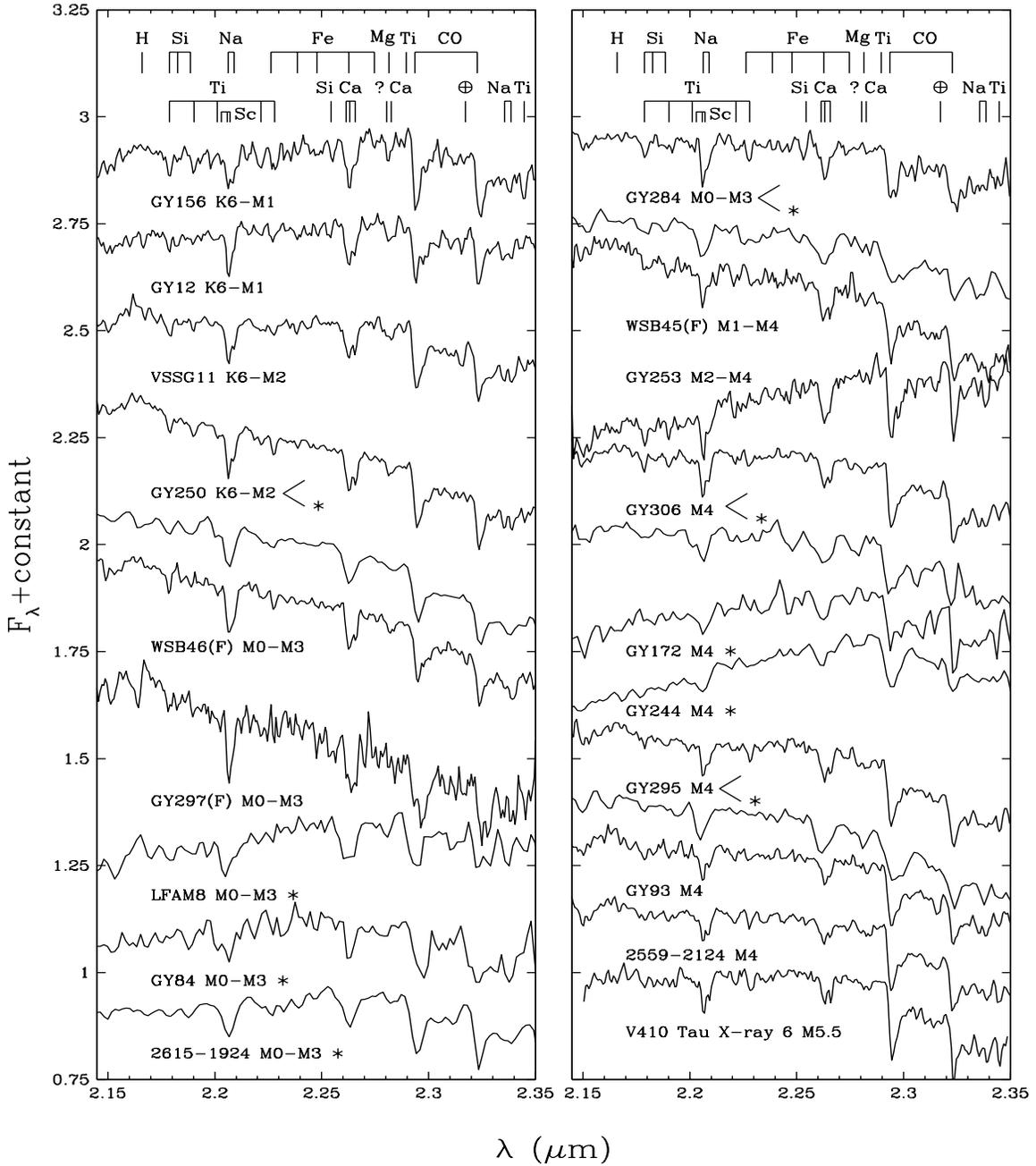}{6.5in}{0}{80}{70}{-245}{-20}
\caption[$K$-band spectra of sources in $\rho$~Oph ($r_{K}<0.25$)]{
$K$-band spectra at $R=1200$ of late-type sources exhibiting
no detectable continuum veiling ($r_{K}<0.25$), with additional data at
$R=800$ ($\ast$). Spectra are normalized at 2.2~\micron\ with constant offsets.
Possible foreground stars (F)
are also indicated. The IR spectrum of V410~Tau~X-ray~6 (LR), which
is classified in the optical as M5.5 (Luhman 1999), is 
shown for comparison. The spectra classified as M4 are also fit marginally by 
veiled late-K stars with enhanced CO absorption.
Spectra are normalized at 2.2~\micron\ with constant offsets.
}
\label{fig:oph.irspec2}
\end{figure}

\begin{figure}
\plotfiddle{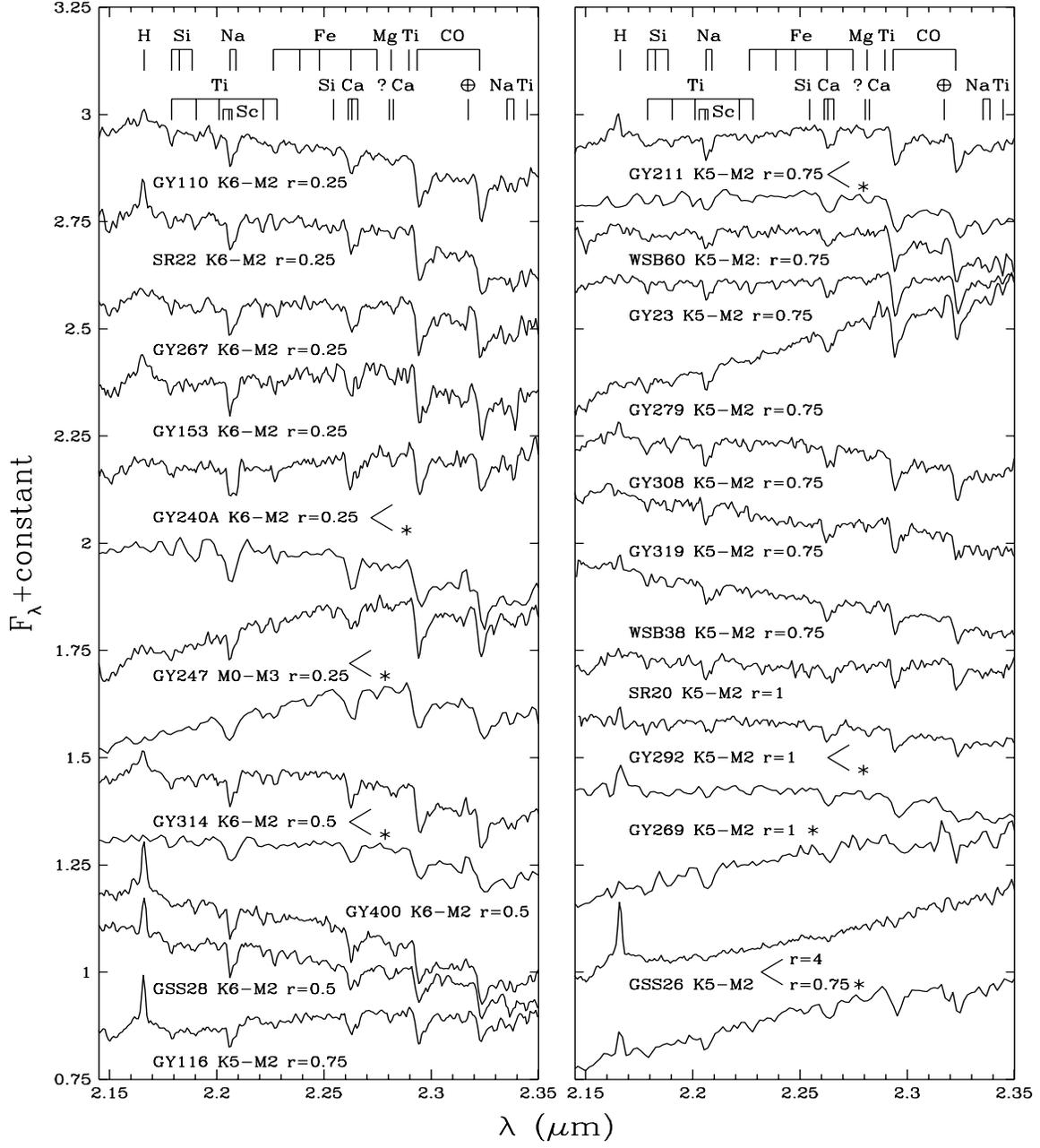}{6.5in}{0}{80}{70}{-245}{0}
\caption[$K$-band spectra of sources in $\rho$~Oph ($0.25\leq r_{K}\leq1$)]{
$K$-band spectra at $R=1200$ of moderately veiled sources 
($0.25\leq r_{K}\leq1$), with additional data at
$R=800$ ($\ast$). Spectra are normalized at 2.2~\micron\ with constant offsets.
}
\label{fig:oph.irspec3}
\end{figure}

In a study of the small stellar aggregate in L1495E in Taurus, 
LR developed a technique of $K$-band spectral classification ($R\sim1000$) 
to derive the spectral types and continuum veilings of young, late-type stars
($\sim1$~Myr, $>$G0).  This method was then used to classify $\sim100$ stars
in IC~348, where the optical and IR spectral types were in reasonable agreement.
The subtleties of $K$-band classification of young stars are discussed by 
LR and LRLL.  For instance, the surface gravity of 
pre-main-sequence stars is intermediate between that of dwarfs and giants,
resulting in weakened Na absorption at both 2.2~\micron\ and 8200~\AA\ in
young M stars relative to M dwarfs (Luhman et al.\ 1998a; Mart{\'\i}n, 
Rebolo, \& Zapatero Osorio 1996).  Comparing the data obtained in IC~348
($R=800$) and that presented here ($R=1200$), we find that
the higher resolution provides more accurate spectral types for G through
M stars while maintaining sensitivity to faint sources. At higher 
resolution we obtain more accurate measurements of Mg at 2.28~\micron, 
as illustrated in the observations of GY240A, GY240B, and GY250 at
$R=800$ and 1200 (Figs.~\ref{fig:oph.irspec1}-\ref{fig:oph.irspec3}).
The strength of Mg relative to Ca is important in the spectral
classification of K through mid-M stars. Mg becomes weaker with later spectral
type and eventually disappears beyond M2. For mid-M and later, 
two other features (Ca and an unidentified line) appear near 2.28~\micron\ and
can be mistaken for Mg when blended together at $R<1000$.
The two lines are resolved at $R\geq1200$, allowing a more accurate 
classification (see V410~Tau~X-ray~6 in Figure~\ref{fig:oph.irspec2}). 

The new spectra are shown in Figs.~\ref{fig:oph.irspec1}-\ref{fig:oph.irspec5} 
in order of spectral type and veiling. Background giants have very distinct 
spectra and are relatively easy to classify in moderate signal-to-noise data. 
A few spectra of lower quality are also shown, where the strong CO absorption
could arise in either background giants or late-type cluster members. The 
spectral slopes, atomic line strengths, and mid-IR emission excesses imply that 
most of these objects are probably the latter. Given the various effects 
of veiling, surface gravity, and temperature on the appearance of these
spectra, degeneracies in the classifications can occur. For instance, the 
sources classified as M4 are also matched by late-K stars with atomic lines 
weakened by veiling and CO enhanced by low gravity.  After comparison of 
IR and optical types in IC~348, we find that generally the simplest IR 
classification is the appropriate one, rather than a spectral type that
requires simultaneous veiling and strengthening of CO. 
Previous optical and IR spectral types listed in Table~1
agree well with our measurements. Exceptions include SR20 and GY410, which were
previously classified as G0: (Bouvier \& Appenzeller 1992)
and M6 (Greene \& Meyer 1995), while we find very good fits with K5-M2 and
K6-M0, respectively. Differences between our classifications and those of
WGM are discussed in the next section.

\subsubsection{Spectral Classification with H$_2$O Absorption}
\label{sec:oph.steam}

In stars later than M5, the atomic absorption lines weaken and are only useful
in classification of high signal-to-noise spectra. To derive 
spectral types for faint young low-mass candidates, we must instead rely on 
absorption in H$_2$O (1.7-2.15~$\mu$m, $\gtrsim2.3$~$\mu$m) and CO 
($>2.29$~$\mu$m).  Because these features are broad and become stronger with
cooler temperatures, lower spectral resolution ($R<500$) is adequate for
measuring spectral types. 
The CO absorption bands of young stars generally appear to arise 
from photospheres rather than disks (Casali \& Eiroa 1996; Greene \& Lada 1997)
and CO and steam should form in the same region of a disk, thus we expect the
measured steam absorption to be dominated by stellar photospheres. 

Detectable $K$-band steam absorption begins in early M dwarfs, causing the
stellar continuum to peak at $\sim2.25$~$\mu$m and slope down to shorter and 
longer wavelengths. These depressions become more pronounced with later M 
dwarfs until M8V (WGM).  Tokunaga \& Kobayashi (1999) found that the 
$K$-band spectra of the field dwarfs vB~10 (M8V)
and DENIS-P J1058.7-1548 (L3V, Kirkpatrick et al.\ 1999) are quite similar 
despite their very different optical spectra. It is likely that the steam
absorption saturates for a large range of temperatures and is thus
a precise indicator of spectral type only for $<$M8V or late L.

Steam absorption can be measured through a reddening-independent index, as 
demonstrated by WGM. 
To estimate spectral types from such an index, the stellar continuum across
the entire $K$-band must be reliably measured, requiring observations of 
telluric standards immediately before and after the targets and the use of the 
same instrument configuration for all observations. Following these guidelines
in observations with FSpec, we find that the spectral slopes are reproducible 
for the same object at different times, while there can be systematic 
differences from instrument to instrument (e.g., FSpec and Kitt Peak IR
Cryogenic Spectrometer). 

For the classification of brown dwarf candidates in $\rho$~Oph, WGM used 
a relation between steam absorption and spectral type observed for M dwarfs.
However, since the surface gravities of these pre-main-sequence objects are
$\sim2$~dex lower than those of M dwarfs, we must investigate whether steam
absorption is sensitive to surface gravity over the regime from dwarfs
to young stars. Thus, we have compared the $K$-band spectra of M dwarfs and 
optically classified young late-type objects.  Spectra covering 2.0-2.4~$\mu$m 
are provided in LR for the young objects V410~X-ray~3 (M6, Strom \& Strom 1994; 
Luhman et al.\ 1998a), V410~X-ray~5a (M5-M5.5, Strom \& Strom 1994; Brice\~{n}o 
et al.\ 1998), and V410~X-ray~6 (M5, Strom \& Strom 1994; M5.5, Luhman 1999). 
When comparing
these data to a spectrum of GL~406 (M6V; same star used to classify Vx3 in the
optical by Luhman et al.\ 1998a), we find that the steam in the young
sources is stronger than in the dwarf counterparts. Therefore, whereas steam 
in dwarfs probably saturates near M8V, saturation may begin at M6 or M7 in 
young sources.  The deep steam absorption present in young objects may seem
contrary to the weak absorption observed at the very low surface gravities
(log~$g=0$-1) of cool giants (Kleinmann \& Hall 1986). However, the synthetic
spectra of Allard et al.\ (1995) do predict the strengthening of steam 
that we observe from M dwarfs (log~$g\sim5.5$) to pre-main-sequence objects 
(log~$g\sim3.5$). 

Using a steam absorption spectral sequence derived for evolved M dwarfs, the 
steam indices would imply spectral types of M6 for Vx5a and M7-M8 for Vx3 and 
Vx6, which are 1-2 subclasses later than their optical spectral types. To 
derive consistent spectral types from the steam absorption, the spectral 
sequence needs to be based on optically classified young stars. With Vx3, 
Vx5a, and Vx6 as spectral standards, we measured spectral types from the 
data of Williams et al.\ (1995) (GY5, GY10, GY37, GY202, 162618-242611) and 
for the coolest objects in our new sample (GY29, GY59). As seen
in Table~1, these spectral types are systematically
earlier than those reported by WGM.  Note that the M6-M7 spectral types may be
upper limits because of the onset of saturation in steam. We classified an
additional star in 
the WGM sample, GY84, with the atomic absorption lines in the manner discussed 
in the previous section and used by LR and LRLL. We find a spectral type of 
early M for this object, in contrast to the M6 classification of WGM.

With the later WGM spectral types, some of the sources are 
very cool yet luminous and fall high above the youngest theoretical isochrones. 
This placement appears to be inconsistent with photometric modeling (CRCTL). 
These sources are moved to the left on the H-R diagram with our classifications 
and have more plausible temperatures given their luminosities and likely ages. 
The revised parameters are more consistent with the photometry of CRCTL.

The above analysis implies that the WGM spectral types are too cool by 
$\sim1$~subclass for M6-M7 and $\sim1$-2~subclasses for $\geq$M8. For WGM
sources that were not in our sample or could not be classified in our data,
we have revised the WGM spectral types accordingly, with the results of CRCTL 
as an added constraint.  For 162618-242416, we adopt M5 as the spectral 
type, which is consistent with the data of both CRCTL and WGM.
Given the apparent systematic error in WGM spectral types, 
the WGM classification of M8 for GY64 is revised to M7, which agrees with CRCTL.
As noted by WGM, the $JHK$ photometry of GY11 indicates heavy continuum veiling,
which would lead to an overestimate of the temperature when steam is used 
in the classification. Therefore, the WGM spectral type of M6.5 probably does 
not need to be adjusted to an earlier type, as suggested by the temperature
estimated by CRCTL, which is equivalent to M7.5.  Finally, three sources 
do not fall in our classified sample or the data of CRCTL.
The relatively early WGM spectral types of GY31 (M5.5) and GY326 (M4) are
adopted while the M8.5 spectral type of GY310 is revised to M7 for use
in our analysis. 

\subsubsection{A Possible FU Ori Object}

On the northeast outskirts of the cloud,
the bright IR source known as IRAS64A (Wilking, Lada, \& Young 1989; Green
et al.\ 1994)
does not appear to be a typical pre-main-sequence star.  The $K$-band spectrum 
exhibits weak atomic lines and strong CO, indicative of a K giant. However, 
the near- and mid-IR excess emission towards this source implies youth and 
cluster membership. As discussed in more detail in \S~\ref{sec:oph.irappend}, 
we find that the data matches that of FU~Ori objects (Hartmann \& Kenyon 1996).
However, this object may also be a background source, such as a 
post-AGB star or Mira variable. 

\subsection{Extinction}
\label{sec:oph.extinction}

\begin{figure}
\plotfiddle{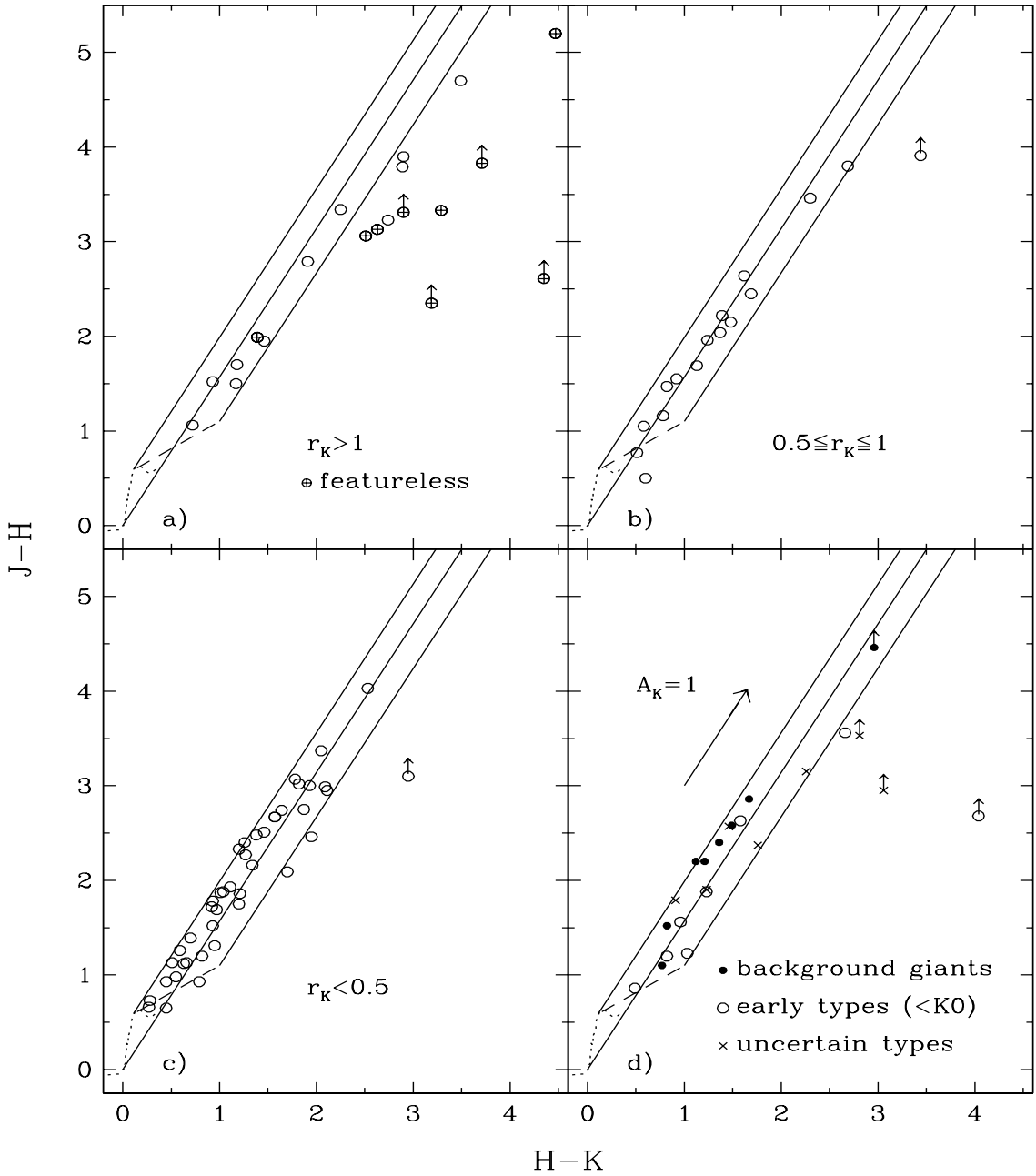}{6.5in}{0}{80}{70}{-245}{0}
\caption[$H-K$ vs.\ $J-H$ for $\rho$~Oph as a function of $r_{K}$]{
$H-K$ vs.\ $J-H$ for the $K$-band spectroscopic sample is given in a), b),
and c) as a function of the continuum veiling, $r_{K}$, 
measured in the spectra. Background giants, early-type stars, and sources
with uncertain late-type classifications are given in d).
The main sequence (dotted line) and
CTTS locus (dashed line) are shown along with the reddening bands of each, where
the reddening vectors were derived from BKLT data by Kenyon et al.\ (1998b).
}
\label{fig:oph.jhhkrk}
\end{figure}

The recent survey by BKLT provides a homogeneous set of coordinates and $JHK$ 
photometry for virtually the entire spectroscopic sample 
(see Table~1). The photometry is in the SQIID 
photometric system, which is similar to CIT (Kenyon, Lada, \& Barsony 1998b).
Photometry and coordinates for ROXs39 and SR22 are taken from Greene et 
al.\ (1994) and references therein.  The measurements of these two 
stars were converted to CIT using transformations from Humphreys, Jones, 
\& Sitko (1984) and Leggett (1992). Since the $\rho$~Oph cloud core
represents a relatively small fraction of the square degree observed by
BKLT, most of their survey region has low reddening and background stars are 
likely to dominate the stellar counts. Under this assumption, Kenyon 
et al.\ (1998b) used the IR color-color diagram to measure an extinction slope
of $E(J-H)/(H-K)=1.57\pm0.03$. We have spectroscopically identified several
background stars, whose colors in Figure~\ref{fig:oph.jhhkrk} are consistent
with this reddening vector. For dereddening individual sources, we used the 
extinction law of Rieke \& Lebofsky (1985). Although it differs from the slope 
found by BKLT, the difference is minor and the slope of BKLT alone cannot be 
interpreted in terms of individual color reddening coefficients.  

Figure~\ref{fig:oph.jhhkrk} shows 
$H-K$ versus $J-H$ diagrams where the main sequence and the
locus of classical T~Tauri stars (CTTS) (Meyer et al.\ 1997) are plotted 
with their respective reddening bands. In the first three panels
we present the spectroscopic sample as a function of $r_{K}$.  The sources 
behave as expected. Those 
with heavily veiled spectra and large $H-K$ excesses fall on the reddened CTTS 
locus or to the right of it. Moderately veiled objects tend to fall between the 
reddened main sequence and the extension provided by the CTTS locus. Objects 
with no detectable veiling and normal colors are consistent with reddened main 
sequence stars. This behavior is similar to the trend observed
by Casali \& Matthews (1992) and Casali \& Eiroa (1996) in spectra of CO 
absorption in a smaller sample of young stars. 
The scatter in this correlation, where a few sources with large $r_{K}$ have 
small excesses in $H-K$, or vice versa, can be explained by variability in the 
excess continuum emission (mostly at $K$). Such variability is evident in the 
data for GSS26 and GY51 (see \S~\ref{sec:oph.nature}.)

Extinctions were estimated for stars earlier than G0 by dereddening to
the main sequence value of $J-H$ for a given spectral type. Reddenings for 
late-type stars are often also estimated by assuming 
main sequence values for the intrinsic colors. 
In an older cluster such as IC~348 (0.5-10~Myr), since much of the population
consists of evolved Class~III sources that show little evidence for disks,
this approach for low-mass stars is reasonable.  On the other hand,
in a very young cluster such as $\rho$~Oph ($<1$~Myr), many of the stars
are Class~I through Class~II with IR colors affected substantially
by emission from circumstellar disks and envelopes.  

\begin{figure}
\plotfiddle{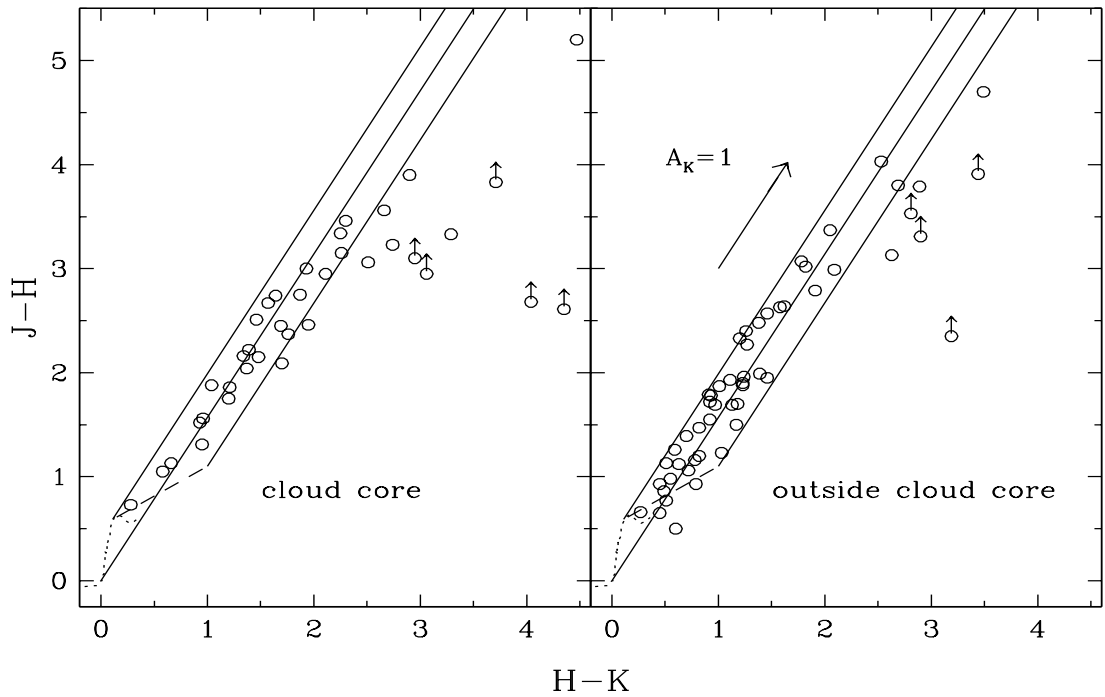}{6.5in}{0}{80}{70}{-245}{0}
\caption[$H-K$ vs.\ $J-H$ for $\rho$~Oph as a function of position]{
$H-K$ vs.\ $J-H$ for the $K$-band spectroscopic sample within and outside the
cloud core, as defined by dashed boundary in Figure~\ref{fig:oph.map}.  
Background 
and foreground stars have been omitted. The main sequence (dotted line) and
CTTS locus (dashed line) are shown along with the reddening bands of each, where
the reddening vectors were derived from BKLT data by Kenyon et al.\ (1998b).
}
\label{fig:oph.jhhkinout}
\end{figure}

By using $E(R-I)$ to deredden the $J-H$ and $H-K$ colors, Meyer et al.\ (1997) 
derived a locus of intrinsic colors for CTTS in Taurus, which 
are equivalent to Class~II sources. The CTTS locus is shown as the dashed line 
in Figs.~\ref{fig:oph.jhhkrk} and \ref{fig:oph.jhhkinout} and has been 
reproduced by models of star-disk systems (Lada \& Adams 1992; Meyer, Calvet,
\& Hillenbrand 1997). 
The CTTS locus of Meyer
et al.\ (1997) was measured for stars of spectral types near M0, where
the origin of the locus falls at the main sequence colors of M0. As 
predicted by Meyer et al.\ (1997) and observed by Luhman (1999) in IC~348,
the locus and its origin are shifted to redder $H-K$ for later M spectral
types. For earlier stars that fall within the reddening band of the 
CTTS locus, extinctions were calculated by dereddening 
the $J-H$ and $H-K$ colors to this locus. Sources to the right of the CTTS 
reddening band in Figure~\ref{fig:oph.jhhkrk} are heavily veiled and have 
uncertain spectral types, consequently they are not plotted in the H-R diagram.

\subsection{Effective Temperatures and Luminosities}
\label{sec:oph.tefflbol}

To convert spectral types and photometry to effective temperatures and
bolometric luminosities, we have adopted the temperature scale and procedures 
discussed in detail by LR and LRLL.  We combined the dereddened $J$ photometry 
with the bolometric corrections compiled by Kenyon \& Hartmann (1995) and a
distance modulus of 6.1 (Whittet 1974) to arrive at the bolometric luminosities.
The source extinctions at $J$, effective temperatures, and bolometric 
luminosities are listed in Table~1, where
the interstellar reddening law of Rieke \& Lebofsky (1985) is used with 
$A_{J}$ in the Johnson-Glass photometric system.

No $J$ photometry is available for two sources in our sample, GY244 and 
GY269.  Since both show strong emission in the mid-IR, the $H-K$ color may be 
contaminated significantly by excess at $K$. Our measurements of $r_{K}$
from the spectra could be used to correct for this effect, but since 
veiling can be variable and
the spectra and colors were not measured simultaneously, we refrain from
this approach. Although we cannot reliably place it on the H-R diagram, 
we will use the spectral type of GY244 and a canonical age to arrive at
an approximate mass that can be added to the IMF for the cloud core in
\S~\ref{sec:oph.imf}.  Such an estimate was not necessary for GY269,
which falls outside of the cloud core. For the B star S1, we refer to the 
calorimetric luminosity derived by Wilking, Lada, \& Young (1989).

\section{The $\rho$~Oph Stellar Population}

\subsection{Cluster Membership}

Classic methods of determining cluster membership (e.g., proper
motions, Li absorption) are unavailable for most of the stars in our
spectroscopic sample. However, we can confidently distinguish cluster 
members from field contaminants -- foreground and background stars --
through other means.  With $K$-band spectra of modest signal-to-noise, 
background giants are easily identified through their distinctive continuum
structure and strong CO absorption. All of the background giants in our sample
appear in the less reddened area surrounding the $\rho$~Oph cloud core, as
indicated in Figure~\ref{fig:oph.map}.  The sample of WGM includes one likely
background giant and two background early-type stars. GY297 
shows little or no reddening in the IR colors and has no other indications of 
youth or cluster membership, such as near- or mid-IR excess emission or 
weakening of Na expected for young M stars (Luhman et al.\ 1998a, 1998b).
It is a probable foreground star and is omitted from further analysis. 
WSB45 and WSB46 are a close pair ($10\arcsec$) that show no obvious reddening
in their IR colors or Li absorption (WSB45, Bouvier \& Appenzeller 1992). 
Strong, variable X-ray emission and weak H$\alpha$ emission (3-6~\AA) have been
detected towards these sources (Montmerle et al.\ 1983), suggesting
they are either pre-main-sequence stars or active field dwarfs. We omit these
stars for our discussion here.  The remaining sources
are unlikely to be background or foreground stars since
they fall above the main sequence and have reddened colors. To examine cluster 
membership at fainter magnitudes beyond the limit of our spectroscopic sample, 
we will use a newly acquired KLF and diagrams of $H-K$ versus $K$ to develop
a model for the background star population shining through the cloud core
(\S~\ref{sec:oph.redmodel}).

\subsection{The H-R Diagram and Star Formation History}
\label{sec:oph.hr}

\begin{figure}
\plotfiddle{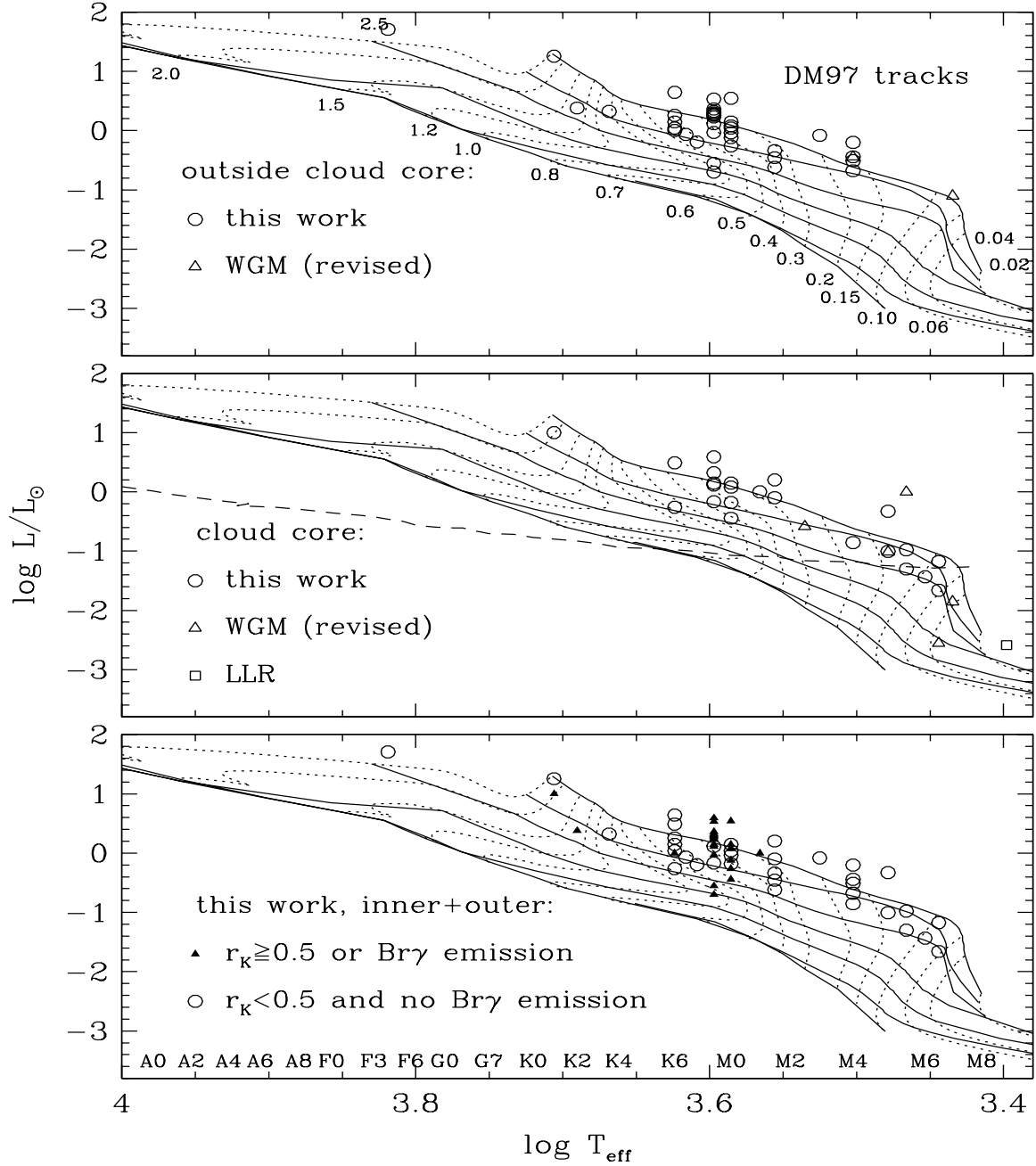}{6.5in}{0}{80}{70}{-245}{-30}
\caption[H-R diagram for $\rho$~Oph]{
The H-R diagram of the $\rho$~Oph star forming region outside (upper panel) 
and within (middle panel) the cloud core, as defined by the dashed boundary 
in Figure~\ref{fig:oph.map}. Low-mass sources observed by WGM and 
Luhman et al.\ (1997) (LLR) are also shown.  The lower panel
compares sources in our sample with and without signs of disk activity.
The theoretical evolutionary tracks of DM97 are given, where the horizontal 
solid lines are isochrones representing ages of 0.3, 1, 3, 10, 30, and 100~Myr
and the main sequence, from top to bottom. 
The dashed line in the H-R diagram for the cloud core represents
$K_{\rm dereddened}=11$, which is the approximate completeness limit of
the spectroscopic sample for $A_K<3$.
}
\label{fig:oph.hr97}
\end{figure}

We use the estimates of $T_{\rm eff}$ and $L_{\rm bol}$ from
\S~\ref{sec:oph.tefflbol} to place the $K$-band spectroscopic sample
on H-R diagrams in Figure~\ref{fig:oph.hr97}. All additional sources 
observed by WGM and 
Luhman et al.\ (1997) are also included. We have omitted objects with uncertain 
spectral types ($<$F8, K-M, featureless) and the foreground and background 
stars.  The diagram of $H-K$ versus $K$ in Figure~\ref{fig:oph.hk} indicates 
that the core spectroscopic sample is complete to $K_{\rm dereddened}=11$
for $H-K<2$ ($A_{K}<3$), corresponding to the dashed line on the
H-R diagram for the core. 

To interpret the H-R diagram of $\rho$~Oph, several sets of low-mass 
evolutionary tracks are available. Luhman (1998), LR, and LRLL made 
comparisons of the D'Antona \& Mazzitelli (1994; DM94) and D'Antona \& 
Mazzitelli (1997; DM97) model predictions with observations of YY~Gem and 
CM~Dra, Pleiades, the main sequence, and globular clusters. Whereas 
calculations of both DM94 and DM97 were used in the analysis of data for IC~348 
to provide continuity with previous studies, we will proceed with only DM97 in 
the following discussion. 

As shown in Figure~\ref{fig:oph.hr97},
the tracks of DM97 imply a median age of $\sim0.3$~Myr for the cloud core, with
a few stars older than 1~Myr, particularly at the lowest masses where
the scatter in apparent ages is largest. A similar range of ages is found in 
the sample surrounding the cloud core, where the median age is slightly older. 
A trend towards older ages with increasing distance from the cloud core is more 
obvious when these results are compared to the H-R diagram plotted by Greene \& 
Meyer (1995) for optically visible
stars observed across three square degrees (Bouvier \& Appenzeller 1992).
Greene \& Meyer also derived an H-R diagram for a sample observed with
IR spectroscopy, composed of 5 stars from the core and 14 stars from the 
outer cluster.  Using DM94, they found a distribution of ages 
very similar to that of the cloud core in Figure~\ref{fig:oph.hr97},
although the study was slightly deficient in stars older than 1~Myr 
relative to the work presented here. Considering the uncertainties in
model predictions and luminosity estimates at such early stages of evolution,
absolute ages for these individual stars are not very meaningful.  However, 
when the same set of evolutionary tracks is used from cluster to cluster, it 
is clear that $\rho$~Oph is one of the youngest nearby stellar populations.

\subsection{The Initial Mass Function}

\subsubsection{The Spectroscopic Sample}
\label{sec:oph.imf}

In conjunction with the evolutionary tracks of DM97, we can use $T_{\rm eff}$ 
and $L_{\rm bol}$ to estimate masses for individual sources in $\rho$~Oph
and construct a cluster IMF. Since the cloud core provides an excellent 
screen against background stars, cluster membership and completeness are
more readily addressed and we thus restrict the analysis of the mass 
function to sources within this region (see Figure~\ref{fig:oph.map}). 
Our spectroscopic sample contains 39 objects that fall within the cloud core.
Seven sources (GY6, GY20B, GY205, GY254, GY265, GY214, GY227) exhibit heavily 
veiled, featureless spectra and are excluded from this calculation of the IMF. 
Photometry for the companion GY20B is not available, hence it is absent from
the figures. Three sources in the cloud core, GY192, 162712-243449, and GY192, 
had relatively uncertain late-type 
classifications and were not plotted in the H-R diagrams. They
are indicated by circled dots in Figs.~\ref{fig:oph.hk} and \ref{fig:oph.klf}.
GY192 has strong emission in {\it IRAS} and 
$K-L\arcmin$ measurements and moderate excess in the $J-H$ and $H-K$ colors.
We find that the spectral type ($>$M3) and luminosity
estimate are consistent with the age of the cluster, implying a mass of
$\sim0.2$~$M_\odot$. Strong excess emission in $H-K$ and $K-L\arcmin$ and 
weak $K$-band features are found in $162712-243449$.  The spectrum of GY128 
is very noisy and quantitative measurements of its features are not possible. 
It also exhibits a flat-spectrum SED and large $K-L\arcmin$ excess. Since 
$162712-243449$ and GY128 cannot be classified and appear to be high-excess,
heavily veiled objects similar to the seven featureless sources 
mentioned previously, they are omitted from the IMF.
Two sources in our sample have uncertain reddenings and luminosities
due to strong IR excess emission (GY51) and a lack of $J$ photometry (GY244).
By assuming these objects have ages comparable to the rest of
the cloud core, we combined their spectral types with the evolutionary 
tracks to arrive at mass estimates.  
Because of uncertainties in the spectral types 
and subsequent mass estimates for two of the three early-type stars, the two 
highest mass bins are given widths of $\Delta{\rm log}~M=0.4$. For
GY182/WL16 we use the $H$-band spectral type and a rough luminosity 
estimate to place it in the mass bin centered at ${\rm log}~M=0.35$ 
(1.4-3.5~$M_\odot$) (Biscaya et al.\ 1999).  For the remaining 23 stars from 
our sample and 6 stars from WGM and Luhman et al.\ (1997) shown in the 
H-R diagram of the cloud core, individual masses were estimated with
the evolutionary tracks.  The two lowest mass bins were also doubled in 
width to compensate for the greater uncertainties in the evolutionary 
tracks and temperature scales in this regime. 

\begin{figure}
\plotfiddle{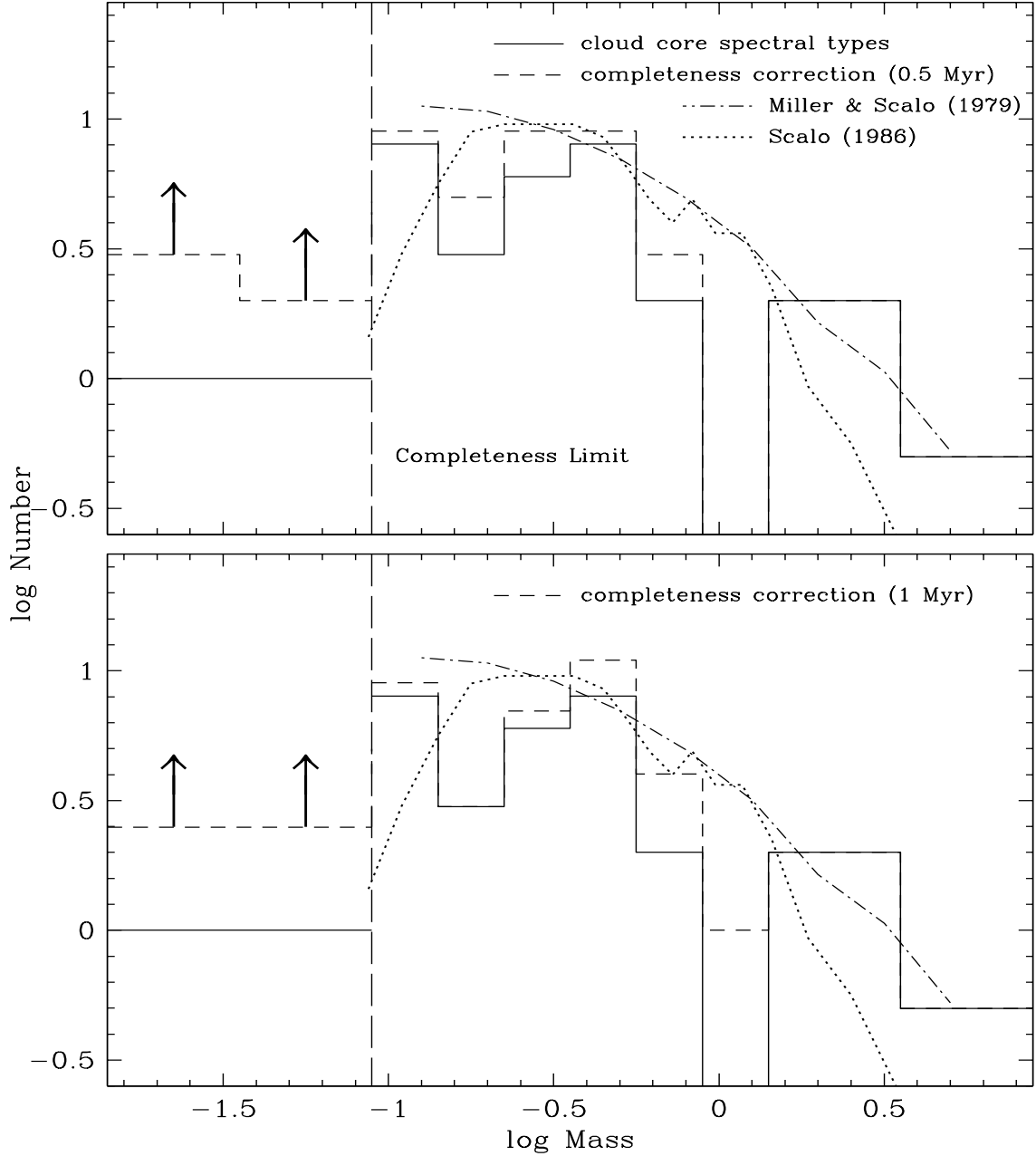}{6.5in}{0}{80}{70}{-245}{-30}
\caption[IMF for $\rho$~Oph]{
The solid histogram is the IMF derived from the cloud core spectroscopic sample 
with the evolutionary tracks of DM97. All sources with featureless spectra 
are excluded from the IMF. The dashed histogram 
includes sources that 1) have uncertain late-type spectral types, 2) are 
identified as likely cluster members in {\it ISO} observations by CRCTL, or 3) 
lack spectra and are an excess above the background population in the KLF in 
Figure~\ref{fig:oph.klf}. The completeness limit is indicated by the 
vertical dashed line and a lower limit to the substellar IMF is provided 
below this boundary.  For reference, the field star mass functions of 
Miller \& Scalo 
(1979) and Scalo (1986) are given.  The two lowest and two highest mass bins 
were given widths of ${\rm log}~M=0.4$ due to uncertainties in mass estimates.
}
\label{fig:oph.imf}
\end{figure}

The resulting IMF for the cloud core is composed of 36 sources, 
represented by the solid histogram in Figure~\ref{fig:oph.imf}.

\subsubsection{Completeness Correction for $K\leq13$}
\label{sec:oph.complete}

A completeness correction can be applied to the above IMF by 1) developing 
a model for the background stars shining through the cloud core at $K$, 2) 
identifying likely cluster members in the cloud core KLF that lack spectroscopy,
and 3) estimating their masses by combining a canonical cluster age implied
by the H-R diagram with photometry and evolutionary tracks. We have applied 
this approach previously in our studies of L1495E and IC~348 (LR and LRLL).

In the case of $\rho$~Oph, we have obtained deep $K$-band images of the cloud
core ($A_{V}\gtrsim50$), covering the same region surveyed by CRBR (see 
Figure~\ref{fig:oph.map}). 
For the following discussion, we have merged our photometry with that of 
BKLT and CRBR. We use the data of BKLT when both $H$ and $K$
are provided ($H\lesssim15.5$, $K\lesssim15.0$), photometry of CRBR for the
remaining sources detected in that work ($H\lesssim17$, $K\lesssim15.5$), 
and our new data for the faintest stars.  
The resulting set of photometry includes objects found in our survey
and in that of BKLT but not reported by CRBR: $162706-243811$,
$162713-243330$, $162726-244045$, $162648-242836$, and $162622-242254$ (GY12). 
The first four sources are faint companions to brighter objects while
GY12 falls within bright nebulosity, explaining their absence from the
photometry of CRBR. We reject five stars detected by BKLT that do not appear 
in our images or those of CRBR: $162718-243433$, $162720-243820$, 
$162703-243726$, $162719-244156$, and $162719-244122$.  

\begin{figure}
\plotfiddle{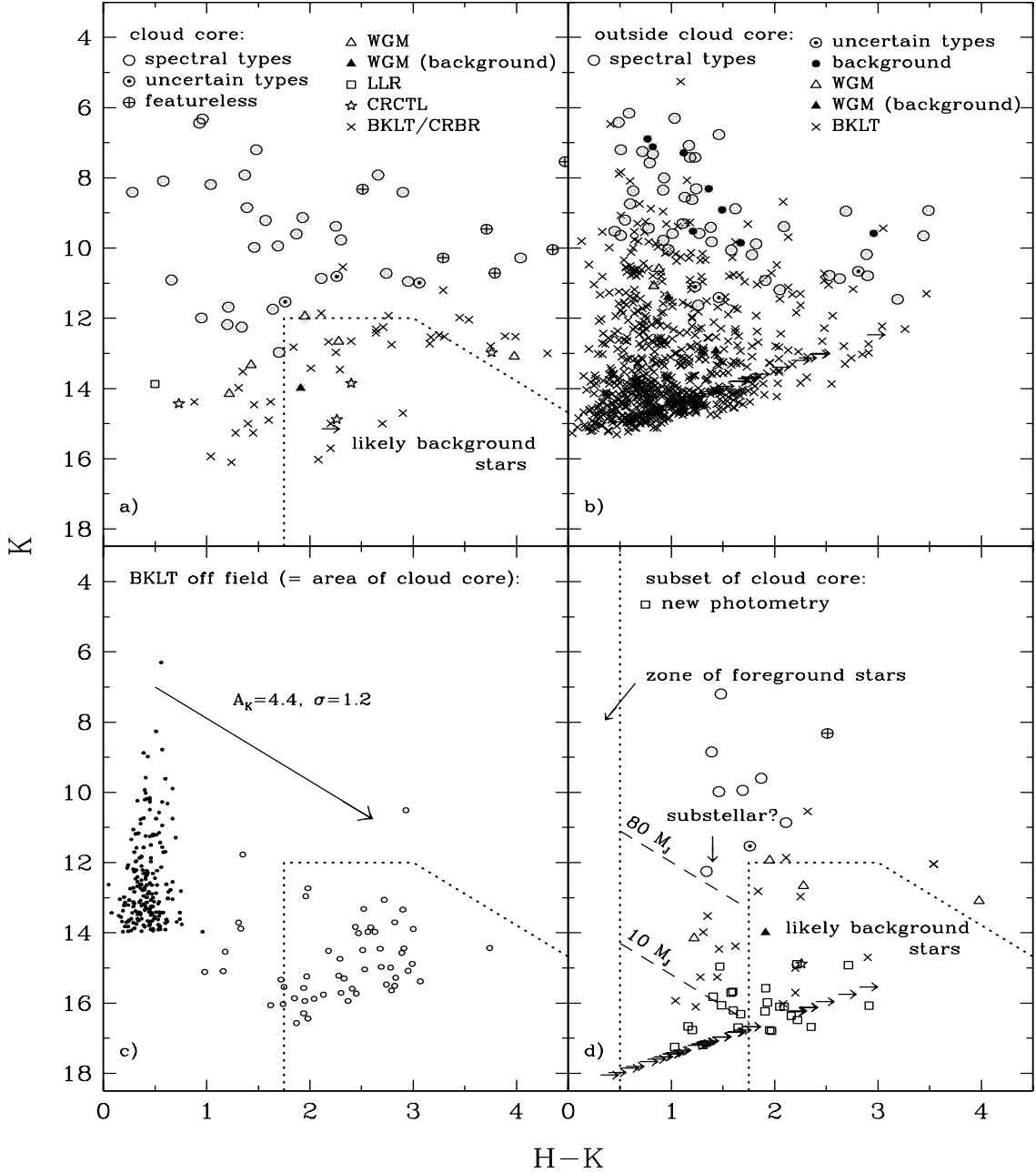}{6.5in}{0}{80}{70}{-245}{-30}
\caption[$H-K$ vs.\ $K$ for $\rho$~Oph]{
$H-K$ vs.\ $K$ for the area a) within the $\rho$~Oph cloud core, as defined
by the dashed boundary in Figure~\ref{fig:oph.map}, b) outside the cloud core 
and within Figure~\ref{fig:oph.map}, c) an off field observed by BKLT equal 
in area to the cloud core, and d) fields in the cloud core outlined by 
dotted lines in Figure~\ref{fig:oph.map} which were observed at $J$, $H$, 
and $K$.  In a) and b), we show sources classified through our $K$-band 
spectroscopy. Additional sources have been observed 
by WGM ($K$-band spectra), Luhman et al.\ (1997) (LLR) (optical 
spectrum), CRCTL ({\it ISO} photometry), and BKLT and CRBR ($JHK$ photometry).  
The KLF (entire cloud core) and the HLF (subset of core) are consistent with a 
gaussian reddening distribution ($A_{K}\sim4.4$, $\sigma\sim1.2$) towards the 
background stars behind the cloud core, which is simulated in c) by reddening
the off-field population. Consequently, the region enclosed by the dotted lines
is where most background stars are expected to appear in a) and d).
The dashed lines in d) are the theoretical $K$ magnitudes of substellar objects
at 1~Myr (Burrows et al.\ 1997).
}
\label{fig:oph.hk}
\end{figure}

\begin{figure}
\plotfiddle{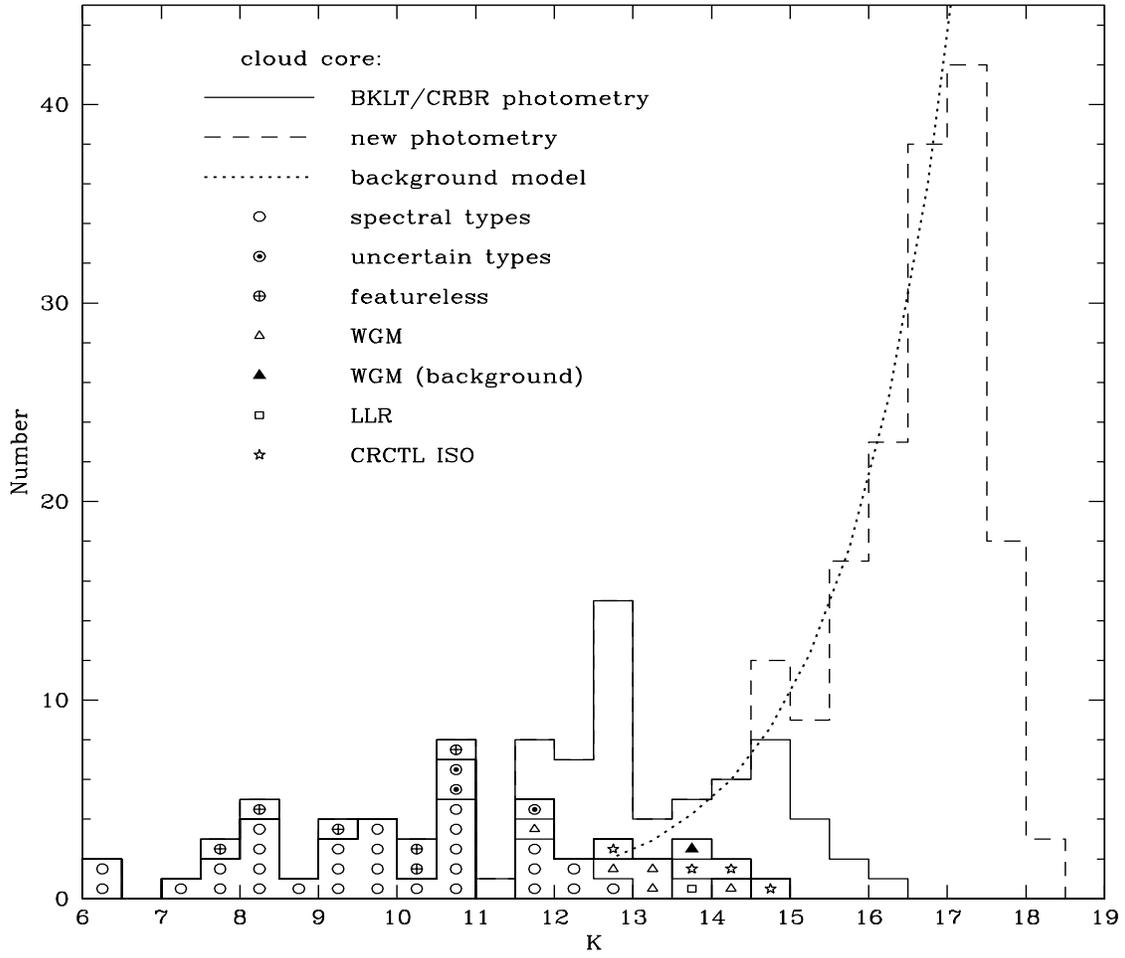}{6.5in}{0}{80}{70}{-245}{0}
\caption[$K$-band luminosity function for $\rho$~Oph]{
The $K$-band luminosity function for the cloud core of $\rho$~Oph. 
The dotted line is a model for the distribution of background stars.
}
\label{fig:oph.klf}
\end{figure}

The cloud core KLF for the combined BKLT and CRBR data is 
given in Figure~\ref{fig:oph.klf}, in addition to the deeper extension provided 
by our data.  It is evident from the rapid rise in counts at $K>16$ that the
new KLF reaches the background star population behind portions of the cloud 
core.  To simulate this field population, we selected all BKLT sources
within a $25\arcmin\times10\arcmin$ area centered at 
$\alpha=16^{\rm h}24^{\rm m}21\fs30$, $\delta=-24\arcdeg42\arcmin00\arcsec$
(1950), which roughly equals the size of the cloud core. This position likely 
contains only relatively unreddened field stars, because it is 
near the outskirts of the BKLT survey where the molecular column density
associated with $\rho$~Oph is quite low.  This off-field KLF must be reddened 
by $A_{K}\sim3.75$ to fit the background star KLF ($K>16$) observed towards
the core. Such an average extinction is also consistent with the lack of 
background stars at $K<12$ in the spectroscopic sample for the cloud core.
The off-field KLF is fit well with a 
function of the form $N(m)dm\propto10^{\alpha m}dm$, where $\alpha=0.31$, 
which is consistent with the Galactic field distribution observed by Wainscoat 
et al.\ (1992). When a KLF of this form is reddened, Comer\'{o}n, 
Rieke, \& Neuh\"{a}user (1999) found that the exponential form and value of
$\alpha$ are preserved regardless of variable extinction or clumps. Thus, we
show this fit to the background stars within the cloud core KLF in
Figure~\ref{fig:oph.klf}.  In a diagram of $H-K$ versus $K$ in the third panel 
of Figure~\ref{fig:oph.hk}, we have reddened the off field by the best fit 
extinction derived in \S~\ref{sec:oph.redmodel}, illustrating where background 
stars should appear in the photometry.

Nine stars showing either little or no features and high 
excess emission were excluded from the IMF determination in \S~\ref{sec:oph.imf}
since we could not constrain their spectral types or place
them on the H-R diagram.  Similarly, we do not
attempt to add the high-excess, featureless objects to the IMF in the following
completeness correction. Since many of the stars with no spectra at
$K<13$ have $L\arcmin$-band measurements, we will reject objects with 
$K-L\arcmin\gtrsim2$, which is typical of Class~I sources. The mass function 
derived from this analysis will
be based on the assumption that these objects are not biased towards 
one mass regime.  This assumption is consistent with recent optical 
spectroscopy of Class~I objects in Taurus, which exhibit a range of spectral
types similar to that found among T~Tauri stars (Kenyon et al.\ 1998a).

CRCTL detected mid-IR emission in {\it ISO} observations of low-mass candidates 
in the cloud core, several of which have been classified by WGM and Luhman et 
al.\ (1997) and added to the IMF in \S~\ref{sec:oph.imf}. The four remaining
sources that lack spectroscopy show weak mid-IR excess emission and can be
added to the IMF by the photometric modeling described below. 
These objects ($162623-242603$, $162653-243236$, $162710-242913$, CRBR33)
are represented by the open stars in the Figs.~\ref{fig:oph.hk} and
\ref{fig:oph.klf}.

The simulated background KLF predicts that only three stars in the $K=12$-13 
magnitude range are background. The cloud core KLF contains a number of sources 
in excess over the background model for $K=11.5$-13. In addition, many of 
the objects between $K=12$-13 have $H-K>3$, much larger than expected for 
background stars, as shown in Figure~\ref{fig:oph.hk}.
We therefore assume that all the objects with $K=12$-13 are members of the
embedded population, except for three we rejected that have plausible colors
to be candidate background objects.  Within these 19 remaining objects, 9
show evidence for strong IR excess emission in $H-K$ or $K-L\arcmin$ and are
excluded from the IMF modeling. 

The remaining ten sources
are added to the IMF as a completeness correction in the following manner.
As in \S~\ref{sec:oph.extinction}, extinctions are calculated by dereddening
the $J-H$ and $H-K$ colors to the CTTS locus. When $J$ photometry is not
available, an intrinsic color of $H-K\sim1$ is assumed, which is 
the maximum expected on the CTTS locus.
Given the star formation history implied by the DM97 tracks in 
Figure~\ref{fig:oph.hr97}, we assume canonical ages of 0.5 and 1~Myr for 
these 10 objects and convert the dereddened $J$ (or $H$) magnitudes to
masses for each age. After adding these photometrically
derived masses to the IMF for the spectroscopic sample, we arrive at the
distributions indicated by dashed lines in Figure~\ref{fig:oph.imf}. 
The completeness corrections for ages of 0.5 and 
1~Myr are very similar.  With the exception of the sources with large IR excess 
emission and featureless spectra omitted from the IMF, the final mass 
function for the cloud core is complete to $\sim0.08$~$M_{\odot}$.

The derived IMF is only the primary star mass function. Although a large number 
of binary systems have been resolved (separation $>0\farcs005$) in lunar 
occultation observations of $\rho$~Oph (Simon et al.\ 1995), our derivations 
include only companions that are detected
in the images of BKLT and CRBR (separation $\gtrsim1\arcsec$). Our IMF can be 
compared directly to results in other clusters, which generally also apply only 
to primary stars. As discussed by LRLL, under a broad range of
assumptions, the inclusion of binary companions in the 
IMF produces a single star mass function with a low-mass slope of
$\sim0.2$-0.7 greater than that of the primary star mass function.

\subsubsection{General Behavior of the IMF}
\label{sec:oph.behavior}

At masses higher than $\sim0.4$~$M_{\odot}$, the DM97 
IMF for $\rho$~Oph matches that of Miller
\& Scalo (1979), while falling less steeply than that of Scalo (1986).
After the peak at $\sim0.4$~$M_{\odot}$, the IMF 
slowly declines to the hydrogen burning limit with a slope of $\sim-0.5$
in logarithmic units (where Salpeter is $+1.35$).  The lower limits for counts 
of young brown dwarfs suggest that the slope is not significantly less 
than $-0.5$ from 0.02 to 0.4 $M_{\odot}$. Allowing for the errors and 
incompleteness below the hydrogen burning limit, these results 
are consistent with the logarithmically flat IMFs between 0.05 and 
1~$M_{\odot}$ estimated by CRBR and Strom et al.\ (1995) in IR 
luminosity function modeling of $\rho$~Oph. 


Using the tracks of Baraffe et al.\ (1997), Bouvier et 
al.\ (1998) have recently reported a slope of $-0.4$ for the low-mass IMF 
in the Pleiades, which agrees with our results in $\rho$~Oph and IC~348. 
LR, LRLL, and Scalo (1998) compared the IMFs measured near 0.1~$M_{\odot}$ 
in photometric and spectroscopic studies of young, open, and globular clusters.
The new results for $\rho$~Oph and the Pleiades further support their
suggestion that these various regions do not show substantial
variation in the low-mass IMF. 

In comparison, recent discoveries of low-mass stars and brown dwarfs by 2MASS 
imply a present day field mass function that rises more rapidly with a slope 
of $+0.3$ (Reid et al.\ 1999). In addition, there appear to be significant 
differences in the mass functions derived between 0.5 and 2~$M_{\odot}$
among these studies.  The Pleiades IMF of Bouvier et al.\ (1998) is 
broad and peaks at 0.4~$M_{\odot}$, consistent with our results in IC~348 
and $\rho$~Oph and the IMF of Miller \& Scalo (1979). However, compared with
this IMF, the population in Orion is deficient in stars over
this mass range and is more sharply distributed at low masses,
with a peak at 0.2~$M_{\odot}$ (Hillenbrand 1997). 

Some of these variations may
be due to differences in the techniques and tracks used in the young clusters.
For instance, the Pleiades data were converted to masses with theoretical 
relations of $M_{\rm bol}$-$mass$ ($>0.7$~$M_{\odot}$, DM97) and
$M_{I}$-$mass$ ($<0.7$~$M_{\odot}$, Baraffe et al.), whereas 
spectral types and H-R diagrams were used for Orion (DM94), IC~348 (DM94
and DM97), and $\rho$~Oph (DM97).  However, these explanations may not account
for the behavior of Orion; whether it is different
from other young clusters remains an open question.

\subsubsection{Comparison to the Pre-Stellar Clump Mass Function}
\label{sec:oph.clump}

\begin{figure}
\plotfiddle{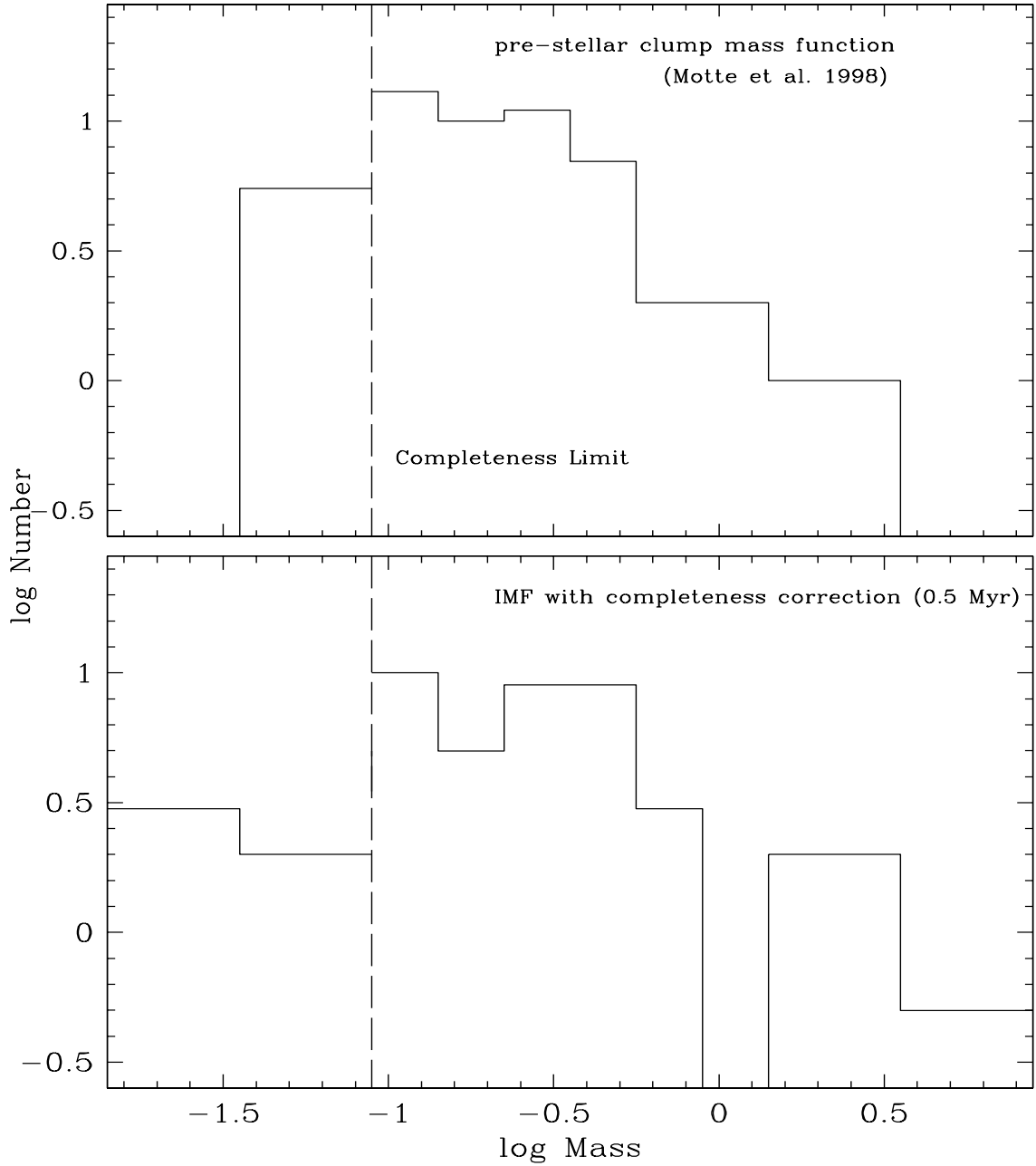}{6.5in}{0}{80}{70}{-245}{0}
\caption[IMF and clump mass function for $\rho$~Oph]{
The pre-stellar clump mass function measured in $\rho$~Oph from millimetter 
continuum
observations (Motte et al.\ 1998) and the cloud core IMF after a completessness 
correction assuming an age of 0.5~Myr (dashed line in lower panel of
Figure~\ref{fig:oph.imf}).  The completeness limits of both studies are near
0.1~$M_\odot$.
}
\label{fig:oph.clumps}
\end{figure}

The proximity and extreme youth of the $\rho$~Oph star forming region make
possible the important comparison of the stellar IMF to the mass function of 
pre-stellar clumps. In deep millimeter continuum observations of $\rho$~Oph,
Motte, Andr\'{e}, \& Neri (1998) have studied emission 
associated with circumstellar material, dense cores, and the ambient cloud.
They detected 58 starless clumps that showed no radio or IR sources and 
appeared to be pre-stellar or protostellar in nature. After estimating
individual clump masses, they constructed a mass function that exhibited
a slope of $\sim1.5$ for $\gtrsim1$~$M_{\odot}$ and $\sim0.5$ for 
$\lesssim1$~$M_{\odot}$. Motte et al.\ noted the similarity to the field 
star mass function (e.g., Miller \& Scalo 1979) and suggested a connection 
between the clump and stellar mass functions. Testi \& Sargent (1998) have 
reached a similar conclusion in a study of Serpens. In this scenario, most of 
the mass in a given clump eventually accretes into a star or multiple system. 
With the new data presented here, we can now directly compare the stellar IMF
for the $\rho$~Oph cloud core to the clump mass function measured by Motte et 
al.\ (1998), as illustrated in Figure~\ref{fig:oph.clumps}.
Within the fairly low number statistics, the two mass functions are quite
similar, providing further support for the notion that clump fragmentation
directly influences the stellar IMF in a site of clustered star formation
such as $\rho$~Oph.

\subsubsection{Using Reddening Models to Test Membership at $K>13$}
\label{sec:oph.redmodel}

To explore the low-mass IMF further requires that we determine cluster 
membership for objects fainter than $K=13$. Figure~\ref{fig:oph.hk} shows 
that simple KLF modeling is inadequate for this goal. 
A number of faint ($K>13$), relatively blue ($H-K<1.75$) objects appear that may
be low-luminosity cluster members rather than background stars. The isolation 
of these objects suggests that detailed modeling of the reddening that combines
$H-K$ colors with the KLF may be more successful in probing for very 
faint cluster members. 

There are discrepancies in the conclusions reached about the reddening in 
different studies. For example, CRBR confined their study to the region 
within the $A_{V}>50$ contour in the CO maps of Wilking \& Lada (1983). Based 
on a detailed discussion of the properties of the background reddened by 
this amount, they concluded that their sample included a number of 
background stars but not in a quantity that would significantly affect their 
results. In contrast, Kenyon et al.\ (1998b) used the $H-K$ versus $J-H$ 
color-color diagram to argue for an extinction only half as great as found 
from the CO maps, and they contend that CRBR may have sufficiently 
underestimated the contamination by background sources to affect their 
conclusions. Neither of these studies makes a quantitative determination of 
the effects of extinction variability on their conclusions. Variations within 
the CO beam would tend to increase the number of background sources 
seen through the cloud above the estimate by CRBR (as they point out). 
Variations will produce an underestimate of the average extinction 
in the color-color plot of Kenyon et al.\ (1998b), unless great care is taken to
avoid selection effects against the redder sources.

With the availability of spectra and {\it ISO} data for many of the faint 
objects in the CRBR study (CRCTL; WGM; this paper), it 
appears indeed that most are bona fide embedded cluster members, in agreement 
with CRBR rather than Kenyon et al.\ (1998b). We still need to reconcile the 
derivation of strong background contamination by Kenyon et al.\ (1998b) and 
their determination of an extinction level less than half of that from CO maps 
and a ``normal" ratio of CO column and $A_{V}$ (Wilking \& Lada 1983). One 
contributing factor is that BKLT and Kenyon et al.\ (1998b) included more area 
outside the $A_{V}=50$ contour in their study than did CRBR. Another 
possibility is that extinction variations may have significantly affected the 
near-IR color distributions.  

To pursue the latter question, we have used a simple model assuming a gaussian 
distribution of extinctions.  We have confined our study to the same area 
studied by CRBR. Most of this area lies within the CO-derived $A_{V}=50$ 
contour. Our model can be used to take a background population and redden it in 
a manner that agrees with the observed properties of the background seen to 
be penetrating the $\rho$~Oph obscuring cloud in our deep $H$ and $K$ images. 
From the background population appearing in off-cloud fields of BKLT, the cloud 
core KLF, and an assumption of uniform extinction, we had deduced 
$A_{K}\sim3.75$ through the cloud (\S~\ref{sec:oph.complete}). However, in deep 
$H$-band images 
of selected regions within the cloud core, we found that the background begins
to emerge at $H-K\sim1.75$. This color is bluer than expected for the average 
level of $A_{K}\sim3.75$, indicating that the extinction is non-uniform, so 
surveys at progressively shorter wavelengths are biased toward progressively 
less obscured members of the background. 

We have two independent measurements to constrain the behavior of the 
extinction. First, by comparing the on-cloud and off-cloud star counts at $K$, 
we have determined the net effective extinction at that wavelength.  Second, 
by imaging deep enough at both $H$ and $K$ to see the background population 
through the cloud, we have a measure of the bias discussed above. The 
interpretation of this bias is relatively simple at these colors, since
plausible background stars all have nearly the same $H-K$ colors. We have 
assumed a uniform value of 0.15, which should be within about $\pm0.1$ 
of the color of most background stars (e.g., Leggett 1992)

We have modeled the extinction of the background assuming that the variations 
are Gaussian around an average value. We have taken a net $A_K=3.75$ as a 
lower limit, to allow for the possibility that the ``off" field we have used 
for comparison with the on-cloud field has some low level of extinction. We 
have assumed an upper limit of the net $A_K<4.05$. From comparing the $H$ 
and $K$ luminosity functions, we find that the color difference at which the 
background emerges is $1.7<H-K<2.2$. 

A range of parameters can fit the observed behavior within these limits. 
The true average extinction could lie from $A_K=3.9$ with a standard 
deviation of 0.6 to $A_K=5.0$ with a standard deviation of 1.45. A 
representative fit is $A_K=4.4$ (i.e., $A_V\sim40$) with a standard 
deviation of 1.2 (see Figure~\ref{fig:oph.hk}), which is our best estimate 
of the true behavior of the 
cloud. Modeling the extinction variations has resulted in an increase by 10 
to 20\% in the estimated average level. However, the possibility that the 
CO measures are slightly biased toward a high estimate is also confirmed.

Our extinction model also allows us to estimate the numbers of background 
sources we might see within our sampled region, as a function of $H-K$ 
(and $K$ magnitude). The proportion is almost independent of $K$ magnitude. 
The models with the largest background contamination are those with high 
average extinction, since the two-parameter fits then require large standard 
deviations for the extinction. 
For these ``worst case" models, $\sim16$\% of the background stars would
have $H-K<1.5$, roughly independent of magnitude, and 
$\sim9$\% would have $H-K<1.25$. Most of the successful models
predicted signficantly lower levels of blue background
star contamination; the values quoted are upper limits.
These predictions are used in the following section
to examine the status (reddened background stars versus low-luminosity 
cluster members) of the very faint ($K>14$), ``blue" ($H-K<1.5$) sources. 

\subsubsection{New Substellar Candidates}
\label{sec:oph.candidates}

Figure~\ref{fig:oph.hk} shows that we cannot achieve complete identifications 
of cloud members from the relatively blue ($H-K<1.5$) colors discussed above; 
at the brighter levels, many bona fide members are redder than this limit. 
However, we can use the predictions to see if we are sampling a portion of the 
very low-mass cloud members. For example, between $K$ of 13 and 16, there are 
25 objects of which 9 are bluer than $H-K=1.5$. Our background model would 
have predicted only 4 objects that blue. The range $K=13$ to 16 corresponds 
to masses from about 0.01 to 0.08~$M_{\odot}$, so the presence of a significant 
number of objects in this range is consistent with our previous arguments that 
the IMF includes a reasonably large number of brown dwarfs at least down to 
0.02~$M_{\odot}$. 

Because of the small numbers of sources included in our deep survey in these
magnitude ranges, this conclusion is not well-established. Our main 
goal is to illustrate a method to extend cloud member identification to very 
low masses. One needs to image the cloud in at least two bands to a sufficient 
depth that the background stellar population is well characterized in both. 
The information on the background star color distribution allows construction 
of a model of the variations in extinction through the cloud. The modeling 
can be quite simple at $H$, $K$, and $L$ because nearly all stars have similar 
intrinsic colors over these bands. Such a model can test whether relatively 
blue sources are likely to be cloud members as opposed to background sources 
shining through relatively thin sight lines of the cloud.    

\subsection{SED Classes and the First Stages of Stellar Evolution}
\label{sec:oph.global}

Through low-resolution ($R=500$) $JHK$ spectroscopy, Greene \& Lada (1996)
examined the near-IR properties of a large number of young stars in $\rho$~Oph
and other star forming regions. A range of IR SED classes was represented in
their sample, including Class~I ($a>0.3$), flat spectrum ($0.3\geq a\geq-0.3$),
II ($-0.3>a\geq-1.6$), and III ($a<-1.6$), where $a$ is the mid-IR spectral
index (Lada 1987; Greene et al.\ 1994). This index measures the slope of an
SED from 2~\micron\ to 10-20~\micron\ and is sensitive to the presence of 
cool circumstellar material. Greene \& Lada observed
a trend of increasing line emission and reddening in addition to
weakening absorption lines from Class~III to Class~I sources, consistent
with the evolutionary interpretation of these observational classes. Our 
results on the relationship between veiling and location on the $H-K$ versus
$J-H$ diagram shown in Figure~\ref{fig:oph.jhhkrk} confirm these arguments.
In addition, we find a decreasing incidence of detectable Br$\gamma$ 
emission with decreasing IR excess, as expected.  

By combining previous near- and mid-IR
photometry with our measured spectral types, continuum veilings, and
emission line strengths, we now investigate in more detail fundamental 
questions regarding the evolution of young stars.  For instance,
are some objects previously identified as Class~I in fact older and 
partially emerged from their natal dust shells, but so heavily reddened by 
the molecular cloud that their SEDs place them in Class~I?
With these data, we will 1) deredden the spectral 
indices and review the SED classifications, 2) 
calculate the timescale for the Class~I protostellar stage
of evolution, 3) use our spectroscopic data to constrain the origin of 
IR excess emission in Class~I and Class~II objects, 4) estimate the 
detectability of absorption features in spectra of Class~I sources, 
5) discuss the nature of the reddening in Class~II sources, and 6) compare the
distribution of SED classes inside and outside of the $\rho$~Oph cloud core.

\subsubsection{Revised SED Classifications}
\label{sec:oph.revised}

To sort the $\rho$~Oph sources into SED classes requires a determination of 
their intrinsic $K(2.2~\micron)-N(10.6~\micron)$ spectral indices. We have 
done so for a sample of all sources with $K<12$ that have
spectral types of K1 or later or exhibit featureless $K$-band spectra.
The multiple systems GY20 and GY240 are not included since the components
are unresolved in the mid-IR data. IRAS64A, the possible background and 
foreground stars, and several sources with uncertain late-type classifications
are also omitted.

Ground-based mid-IR measurements were taken from Elias (1978), Lada \& Wilking 
(1984), Young, Lada, \& Wilking (1986), Wilking, Lada, \& Young (1989), 
and Greene et al.\ (1994). Two objects were classified with the 
{\it ISO} observations of CRCTL.  We avoided {\it IRAS} measurements because 
the beam is large enough that confusion with a fainter and redder object is 
a possibility. We have used these data with the values of $A_{J}$ and $K$ 
in Table~1 to compute the extinction-corrected $K-N$ 
($2.2-10.6$~\micron) spectral indices, $a$(der), given in Table~2.
When available, $a(2.2-20$~\micron) and measurements of continuum veiling 
and Br$\gamma$ emission from the $K$-band spectra are also listed. 

Stars within the reddening bands of the main sequence and CTTS locus should
have reasonably accurate extinction estimates, so for them we have relied on 
the dereddened indices in classifying their SEDs. 
When mid-IR data were available but extinctions could not be computed
due to the lack of $J$-band photometry, the sources were classified by 
considering both $a$(obs) and $r_K$. For sources to the right of 
the CTTS locus, scattered light may dominate and the reddenings are uncertain,
thus we make no attempt at dereddening their spectral indices.
GY6, GY111, GY214, GY254, GY265, GY269, GY274, and GY378 fall in this category
and are likely true Class~I rather than heavily reddened Class~II objects
given their strong veiling and steep SEDs between 2 and 20~\micron. 
GY205, GY224, and GY227 are flat-spectrum sources by their observed indices,
which cannot be dereddened reliably due to
large $H-K$ excess (GY205) and lack of $J$ photometry (GY224, GY227).
The spectra of these three objects are featureless, supporting their 
flat-spectrum status, although the low signal-to-noise provides weak constraints
on the veiling ($r_K>0.5$). 
Because GSS26, GY21, GY51, GY279, and GY315 are within the CTTS reddening band 
or a small distance to the right of it, we can calculate dereddened spectral
indices. Although the dereddened indices for GSS26, GY21, and GY51 indicate
Class~II, they remain close to the flat-spectrum regime, so we cannot 
confidently classify these as either flat or Class~II. Similarly, GY279 
and GY315 are likely Class~II, but we cannot rule out flat-spectrum status
due to variability in GY279 and uncertain extinction in GY315.
No mid-IR photometry is available for GY81, but comparing the spectroscopic
properties to those of the sources we have discussed, it is probably
flat-spectrum or Class~II. Other sources without mid-IR data
are tentatively classified in a similar fashion.  We have no $J$
photometry for GY244, but the lack of veiling or Br$\gamma$ emission implies
little excess at $K$, hence we estimate the extinction from $H-K$. The 
dereddened spectral slope indicates that GY244 may be a heavily reddened 
Class~II object.

For calculating the distribution of sources across SED classes, we divide the 6 
objects labeled as flat/II equally between flat-spectrum and Class~II and
consider flat-spectrum a part of Class~I. 
With the revised classifications, there are 12.5, 26.5, and 33 sources among 
Class~I, Class~II, and Class~III, respectively.  The dereddening has moved
18-30\% (8-13/44) of sources that are Class~I or Class~II by the observed
indices from one class to another, in contrast to the 
suggestion of Greene et al.\ (1994) that less than 10\% of sources should 
change SED classes upon dereddening. These discrepant conclusions may 
be caused by differences in how the dereddened indices were calculated in the
two studies.  Greene et al.\ state that $a$(2-10~$\mu$m)
changes by 0.5 when reddened by $A_V=50$ under the extinction law of 
Rieke \& Lebofsky (1985), whereas we find the change in the index is 1.7.

Selection effects will influence 
these results.  For example, an object of a given mass is expected to fade in 
bolometric luminosity between Class~I and Class~III, favoring 
the detection of the Class~I stage. However, it is also 
possible that a smaller portion of the luminosity emerges in the near-IR
due to the larger extinction in the Class~I phase, favoring the 
detection of the later classes in our $K$-band limited sample.  For instance, 
5/7 sources classified as Class~I or flat-spectrum by Greene et al.\ (1994) 
have $K>12$ and therefore are not included in the statistics we have discussed.
Disks around very young objects where the dust has not started to coagulate 
or dissipate may hide the object from near-IR surveys along some sight 
lines. For example, if the mass of the planets in the solar system were 
distributed uniformly in a disk of 100~AU radius and 3~AU thickness, the 
density would be of order 10$^{-14}$~g~cm$^{-3}$, corresponding to 
$A_{J}\sim1000$
taking typical interstellar parameters. Unresolved binaries will tend to 
favor Class~II designation. In the case of a Class~II source paired with a 
Class~III one, the mid-IR excess would bias the composite to Class~II. A 
Class~II or III source paired with a Class~I one would produce composite 
near-IR characteristics (photospheric absorptions, bluer $K-N$ color) 
indicative of Class~II.
Nonetheless, the ratios of the numbers of sources in the various classes should
give a rough idea of the true distribution among the SED classes.

\subsubsection{Timescale for the Protostellar Stage}

A variety of arguments regarding the very early development 
of a young star assign probable durations to the stages corresponding to 
differing SED characteristics. Specifically, it is thought that the 
Class~I SED type should persist for only $1-2\times10^5$ years (e.g., 
Cohen \& Kuhi 1979; Stahler 1983; Shu, Adams, \& Lizano 1987). However, the 
photometric surveys of CRBR and Greene et al.\ (1994) have 
found Class~I objects in larger numbers than this short duration would suggest. 

However, after dereddening the sources and extracting statistics from
a magnitude-limited sample,
only about 17\% (12.5/72) of them are Class~I. This proportion is 
substantially lower than previous estimates (CRBR; Greene et al.\ 1994).
If we interpret the proportion of Class~I sources as an indication of the 
lifetime of that phase of stellar evolution and assume that star formation has 
occurred at a uniform rate over the last 0.5-1~Myr in the $\rho$~Oph region,
then the lifetime of the embedded stage is $\sim0.075-0.15$~Myr. This estimate
removes the discrepancy with theoretical expectations found in previous 
estimates of the proportion of Class~I sources. 

\subsubsection{Nature of Infrared Excess Emission in Classes I and II}
\label{sec:oph.nature}

In a generally accepted view of the protostellar stage, the Class~I SED
has been modeled as a star surrounded by an accretion 
disk and a thick circumstellar envelope, where the disk is thought
to contribute a substantial portion of the
excess emission at 10~\micron\ (Myers et al.\ 1987). Because of the lack
of disk absorption features in Class~I and II systems, Calvet, Hartmann, 
\& Strom (1997) have suggested that the stellar magnetosphere 
induces a hole in the inner region of the disk, thus preventing the formation
of a disk photosphere.  They propose that a warm infalling envelope dominates 
circumstellar excess at 2~\micron. In a Class~II object where the envelope
is absent and only the disk remains, relatively little $K$-band 
continuum emission ($r_K\lesssim1$) is predicted (Meyer et al.\ 1997). 
We now consider this model in light of the new data we present. 

In May of 1996, GSS26 and GY51 (flat/II) exhibited strong Br$\gamma$ 
emission (2.9$\pm$0.2, 5.0$\pm$0.5~\AA) and veiling ($r_{K}\sim4$, 2), while in 
July of 1994 the Br$\gamma$ emission was weaker (1.7$\pm$0.3, 1.5$\pm$0.5~\AA)
and the excess was lower ($r_{K}\sim0.75$ and 0.5). 
Furthermore, when these spectra are flux calibrated, 
we find that GSS26 and GY51 were $\sim1.2$ and 1.3~mag brighter at $K$ in 
1996 relative to 1994, consistent with the increase in excess continuum
emission implied by $r_{K}$.  Both the excess emission and Br$\gamma$ line flux
in each star brightened by factors of 5-10. If the Br$\gamma$ emission 
arises in columns of material accreting from the disk onto the star
(Muzerolle, Calvet, \& Hartmann 1998), then the simultaneous increase
in the Br$\gamma$ emission and the continuum veiling suggests that the 
excess emission is produced in a region directly associated with the 
accreting material (i.e., the disk) rather than a circumstellar envelope.  

We can also use the behavior of $r_K$ with changing SED class to probe
the origin of the excess emission. 
If an infalling envelope is responsible for the significant continuum 
veiling at 2~\micron\ seen in a given source, strong mid- and far-IR emission
should also be generated, appearing as a Class~I SED. However, we find 
strong veiling, $r_K=1$-4, in several sources lacking this evidence of an
envelope, as shown in Table~2.  Examples include GSS26, GY21, GY51, GY279, 
and GY315, which are possibly Class~II after dereddening. Even without
dereddening, EL24 ($r_K=1$-3), GY129 ($r_K>1$), GY167 ($r_K=1$-3), 
GY168 ($r_K=2$), GY292 ($r_K=1$), and SR4 ($r_K=1.5$-2) are Class~II, yet
they have moderate to heavy veiling. Finally, we measure $r_K=1$ in
GY262 and SR20, which show no sign of an envelope by their Class~III status.
On the other hand, a disk is likely to be present in all of these SED classes.
If the disk is the source of the $K$-band excess emission, then the inner 
hole must be within the region where dust would emit strongly at 
2~\micron\ while large enough to prevent dust destruction and the resulting
formation of a substantial
disk photosphere. The central hole in the model by Greene \& Lada (1996) 
was tuned in this manner and produced veilings of $r_K\lesssim1$ for
a pure re-processing disk and $r_K\lesssim5$ for a luminous disk, although 
this choice of hole size has been subsequently criticized by Calvet et 
al.\ (1997) as nonphysical. 

\subsubsection{Detecting Absorption Features in Class I Sources}

In addition to Class~II and Class~III sources, can photospheric features also 
be detected in extremely red objects designated as Class~I? Greene \& Lada 
(1997) observed absorption features in two sources, GY21 and GY279, 
and demonstrated that the lines likely originate in the photospheres
of the central stars rather than disks. Although these objects were previously
classified as flat-spectrum, they appear to be Class~II when their spectral 
indices are dereddened (see \S~\ref{sec:oph.revised}). With our large set of 
moderate-resolution spectra, we can further examine the
detectability of absorption features in flat-spectrum and Class~I sources.
Featureless $K$-band spectra are found for GY6, GY20B, GY214, GY254, GY265,
GY274, GY378 (Class~I), GY205, GY224, GY227 (flat spectrum), and GY129 
(Class~II).  However, absorption features are detected in the Class~I 
objects GY111 and GY269, comprising about $\sim25$\% of Class~I
sources. Since the signal-to-noise is rather low in some of 
the featureless spectra, higher quality data could reveal absorption features 
in still more of the Class~I stars.

\subsubsection{Nature of the Reddening in Class II Sources}

To test the origin of the obscuration for the highly reddened 
Class~II objects, we have computed the average $A_{J}$ for the Class~II
and Class~III sources in the sample described here. The ratio 
$A_{J}({\rm II})/A_{J}({\rm III})=1.1\pm0.2$; thus there is no excess of 
extinction for Class~II objects. The distribution of extinction values is also 
similar between the classes. Thus, the Class~II and III sources appear to be 
distributed throughout the cloud in a similar fashion, as one would expect. 
This result supports our conclusion that some sources previously considered 
to be of Class~I or II are instead heavily reddened Class~II or III. In 
addition, any 
circumstellar disks around Class~II objects appear to have little influence 
on the extinction -- either the disks are optically thin at $J$, or they are 
thick enough that systems oriented with their disks edge on are so heavily 
obscured that we do not detect them. 

\subsubsection{Spatial Distribution of SED Classes}
\label{sec:oph.spatial}

We now make a comparison of the IR characteristics of the areas separated by the
dashed line in Figure~\ref{fig:oph.map}, defining stellar populations from 
the cloud core and the surrounding region. In the diagrams of $H-K$ versus $J-H$
in Figure~\ref{fig:oph.jhhkinout}, we find that a larger fraction of sources 
within the cloud core have near-IR excess emission. To be more quantitative, 
we designate sources with evidence for actively accreting circumstellar disks as
those exhibiting $W_{\lambda}({\rm Br}\gamma)>1$~\AA\ or $r_{K}\geq0.5$.
In the core of $\rho$~Oph, if we omit the B star and seven sources
with uncertain veiling measurements, the 
frequency of disks is 17/27 (63\%). The frequency is somewhat lower outside of
the cloud core, 23/45 (51\%), after omitting the four earliest stars, two K-M 
stars, IRAS64A, and the foreground and background stars. The difference between 
the two populations is more pronounced when we examine the SEDs.
For sources in the cloud core that have been classified in 
Table~2, the number distribution among Class~I/flat, Class~II, 
and Class~III is 6.5, 9.5, and 10, respectively.
Outside of the cloud core, on the other hand, the SEDs are generally
more evolved, where the distribution is 6, 17, and 23. 

In a study of the older cluster IC~348 (0.5-10~Myr) by LRLL,
$\sim24$\% of sources within the core with ages less than 3~Myr met the
$W_{\lambda}({\rm Br}\gamma)>1$~\AA\ or $r_{K}\geq0.5$ criterion for active
accretion.  The lower fraction of disk sources in the core of IC~348 relative 
to the remainder of cluster was attributed to the accelerated truncation of
disks at higher stellar densities (Herbig 1998; LRLL). In
$\rho$~Oph, since the densities are rather low throughout the region, 
the higher disk frequencies from the core outward are instead likely
due to the birth of stars in the cloud core and their dispersal outward.
Such an age gradient is consistent with the ages implied by the H-R diagram
of the cloud core and surrounding area, as discussed in \S~\ref{sec:oph.hr}.

\section{Conclusion}

We have performed $K$-band spectroscopy towards
$\sim100$ stars within the $\rho$~Oph star forming region. 
We have measured spectral types and continuum veilings 
and have combined this information with previous mid-IR measurements
and new near-IR imaging of the cloud core. Our conclusions are as follows:

\begin{enumerate}

\item
The $K$-band steam absorption, known to strengthen rapidly with from mid-M
to M8V and possibly saturate for later types, 
is also sensitive to surface gravity and differs 
between $\sim1$~Myr and the main sequence for a given spectral
type. Thus, young late-type objects classified in the optical should be used 
as spectroscopic standards when using steam to measure spectral types for 
young, cool objects. 

\item
Using the evolutionary tracks of DM97 to interpret the H-R diagram
for $\rho$~Oph, we estimate stellar ages ranging between 0.1 and 1~Myr, with
a few stars that may be slightly older.

\item
From a deep KLF of the cluster core we develop a reddening model for
the background star population and identify likely cluster members
falling below the limit of the spectroscopic sample.  We use the age from the 
spectroscopic sample to convert their luminosities to masses. After the addition
of these sources, the resulting IMF is complete to $\sim0.08$~$M_\odot$. 
With the tracks of DM97, the IMF for $\rho$~Oph matches that of Miller \& Scalo 
(1979) at masses higher than 0.4~$M_\odot$. The IMF peaks at this mass and 
slowly declines to the hydrogen burning limit with a slope (in logarithmic
units) of $\sim-0.5$, as compared to slopes of $+1.35$, 0, and $-2.6$ for 
Salpeter (1955), Miller \& Scalo (1979), and Scalo (1986), respectively.
The exact shape of the mass function remains
dependent on the theoretical evolutionary tracks and temperature scales. 

\item
Our lower limits on the numbers of substellar objects demonstrate that the 
IMF probably continues a slow decline and does not fall precipitously below 
the hydrogen burning limit, at least down to $\sim0.02~M_\odot$.

\item
The derived IMF is consistent with previous findings that the $\rho$~Oph IMF 
is roughly flat from 0.05 to 1~$M_\odot$. 

\item 
The stellar IMF in $\rho$~Oph is qualitatively similar to the mass function
of pre-stellar clumps, indicating that cloud fragmentation may play a direct
role in determining the IMF.

\item
Very deep imaging at $H$ and $K$ (also at $L$) can be used to construct models 
of the extinction through the cloud and determine background contamination as a 
function of color and magnitude. We have used this method in a limited area 
to identify several faint sources ($K=13$-16) that may
be cluster members by their low reddenings ($H-K<2$), and thus
are extremely low-mass ($10$ to $80$~$M_{\rm J}$) substellar candidates. 

\item
Extinction can significantly alter the true distribution of SED classes. 
The classifications of $\sim25$\% of sources originally classified as 
Class~I-II are changed upon dereddening
of the mid-IR indices ($a$(2-10$\mu$m)). After revising the SED classification
accordingly, $\sim17$\% of the $\rho$~Oph stars are Class~I, implying 
$\sim0.075-0.15$~Myr for the lifetime of this embedded stage.

\item
We detect absorption features in the spectra of $\sim25$\% of Class~I and
flat-spectrum sources, supporting the feasibility of studying 
the photospheres of extremely young protostars. 

\item
In observations of 20 sources separated by two years, we find that the 
continuum veiling and Br$\gamma$ emission change simultaneously in two 
stars. Because the Br$\gamma$ emission likely traces material in
the accretion columns between the disk and the star, this behavior suggests
the disk as the origin of the 2~\micron\ excess rather than an infalling
envelope. The presence of significant veiling ($r_K=1$-4) in the spectra 
of several stars from flat spectrum to Class~III also indicate that a 
disk, or at least something other than an envelope, is capable of producing 
large amounts of continuous 2~\micron\ emission.

\item
The IR properties of the sample imply a slightly more evolved population 
distributed around the cloud core, possibly due to the birth of stars within 
the core and their subsequent dispersal, which is also consistent with the 
ages implied by the H-R diagrams of the cloud core and surrounding region.

\end{enumerate}

\acknowledgements
We thank M. Rieke for helping to obtain the IR spectra and M. Meyer and I. N.
Reid for providing results prior to publication.  Detailed comments on 
the manuscript by C. Lada and M. Meyer were greatly appreciated.
We are grateful to F. Allard, I. Baraffe, A. Burrows, and F. D'Antona
for providing their most recent calculations and useful advice. 
This work was supported by NASA grant NAGW-4083 under the Origins of 
Solar Systems program.

\appendix

\section{Notes on Individual Sources}
\label{sec:oph.irappend}

\begin{figure}
\plotfiddle{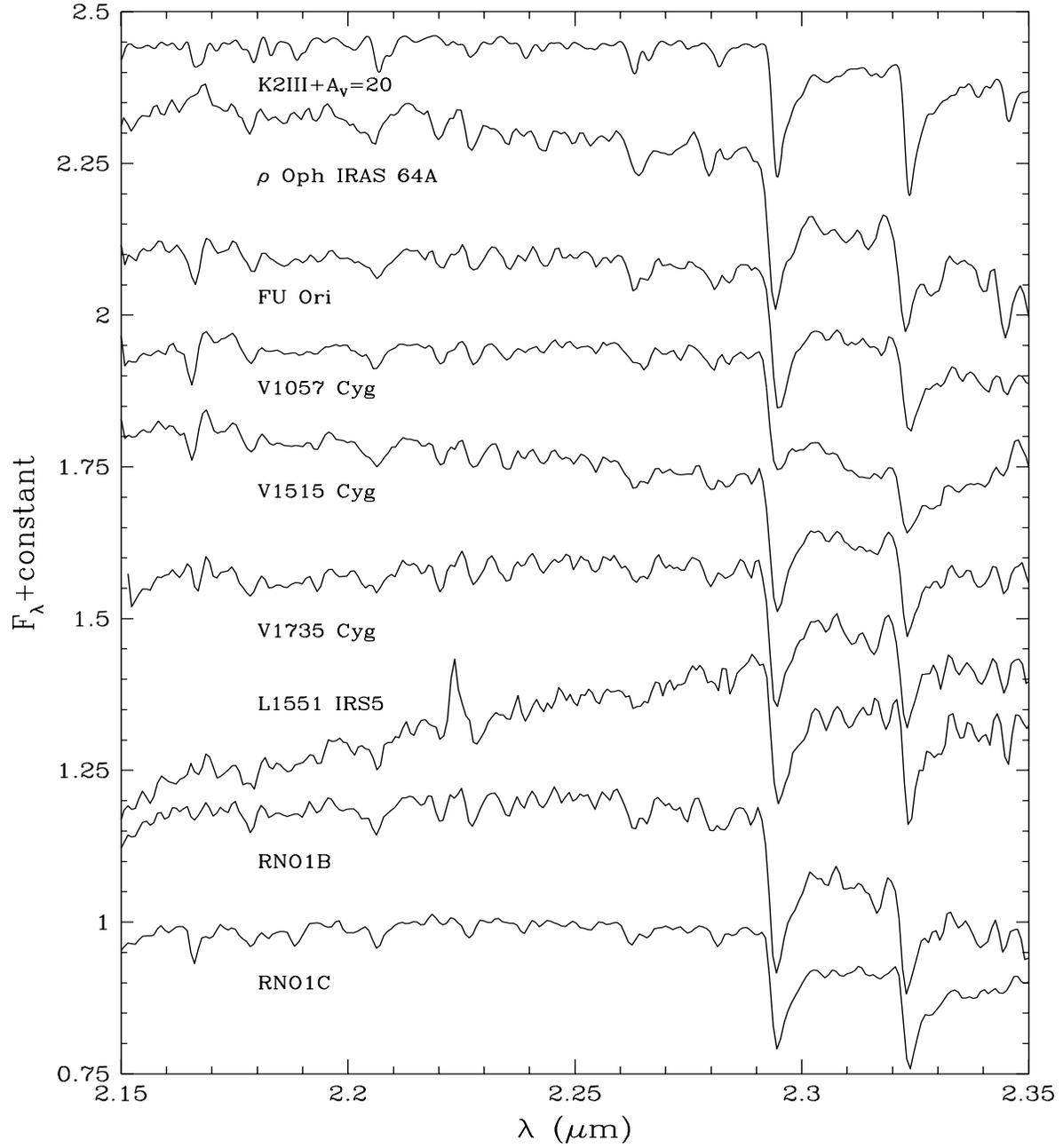}{6.5in}{0}{80}{70}{-245}{0}
\caption[$K$-band spectra of IRAS64A and FU~Ori objects]{
The $K$-band spectrum of IRAS64A is compared to data for several
FU~Ori objects and a typical field K giant (reddened by $A_{V}=20$.)
Spectra are normalized at 2.2~\micron\ with constant offsets.
}
\label{fig:oph.fu}
\end{figure}

With the exception of RNO1C, which is similar to the data for the
K2III $\kappa$~Oph 
(Kleinmann \& Hall 1986), the $K$-band spectra of FU~Ori objects are all 
similar and show characteristics distinct from normal stars, as shown 
in Figure~\ref{fig:oph.fu}. Some of this difference is likely due to 
broadening of lines from disk rotation (Greene \& Lada 1997).  The FU~Ori 
spectra also show two features on either side of where Mg appears in the 
giant (possibly the two lines seen in late M dwarfs) and unidentified 
absorption at 2.33~\micron. All of these characteristics are found in the 
spectrum of IRAS64A.  While the FU~Ori objects have strong CO absorption 
and late-type IR spectra, they tend to appear as F or G giants in the optical. 
This discrepancy is reconciled by attributing the CO to absorption in 
low-gravity disk atmospheres. However, we find that the optical spectrum 
(6000-9000~\AA) of IRAS64A is identical to the M8-M9III spectra presented 
by Kirkpatrick, Henry, \& Irwin (1997).  High-resolution spectroscopy on 
the CO band heads is necessary to determine whether a circumstellar disk 
is the origin of these peculiar optical and IR spectra.

\begin{figure}
\plotfiddle{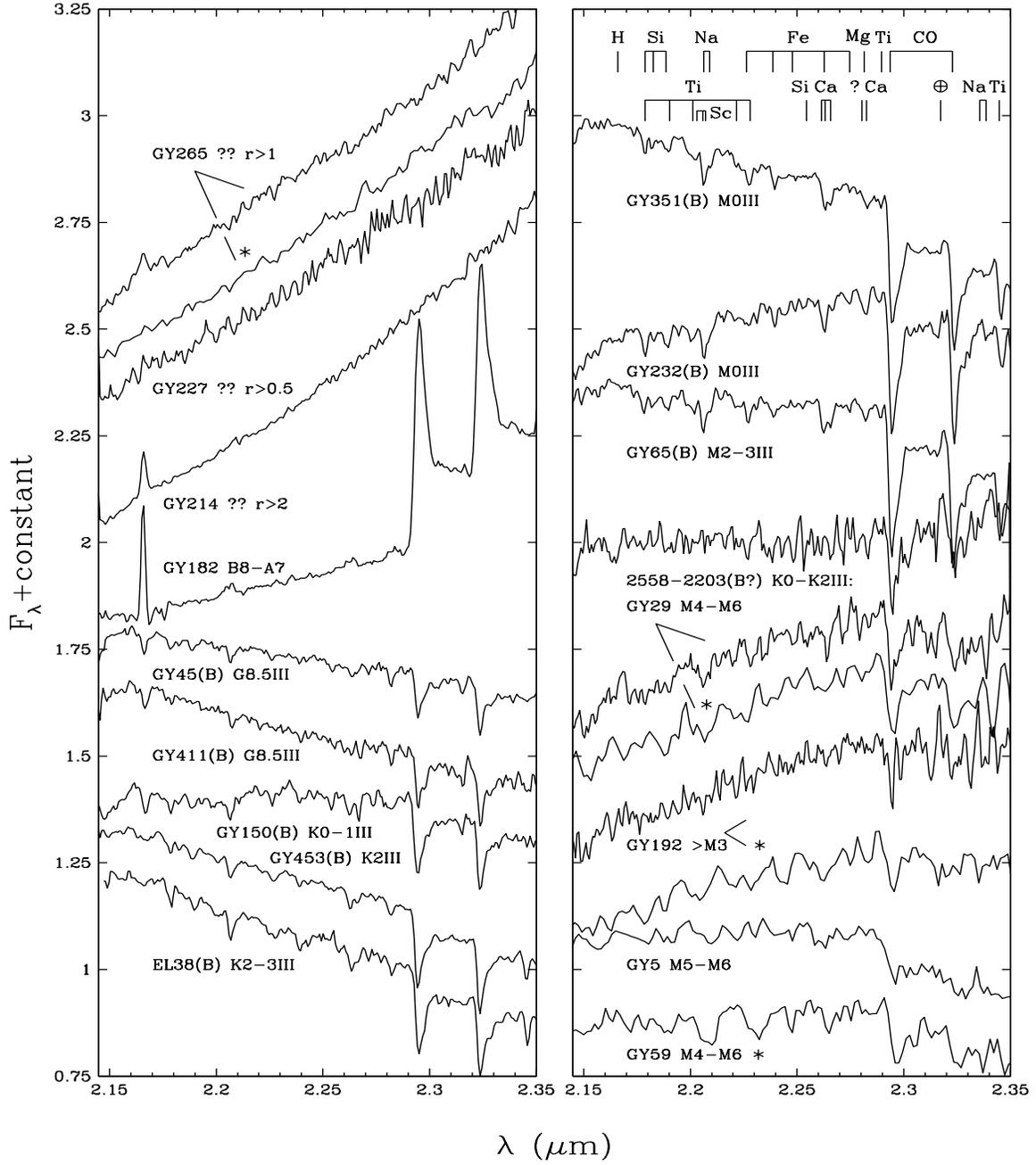}{6.5in}{0}{80}{70}{-245}{0}
\caption[$K$-band spectra of sources in $\rho$~Oph ($r_{K}\geq1$, III, K-M)]{
$K$-band spectra at $R=1200$ of heavily veiled sources ($r_{K}\geq1$),
background giants (B), and objects with uncertain late-type 
classifications, with additional data at $R=800$ ($\ast$).
Spectra are normalized at 2.2~\micron\ with constant offsets.
}
\label{fig:oph.irspec5}
\end{figure}

Another unique source, GY182/WL16, is well-known for its
strong CO band head emission, modeled successfully in terms of an inversion
layer in a circumstellar disk (Carr et al.\ 1993; Najita et al.\ 1996).
As seen in Figure~\ref{fig:oph.irspec5}, no photospheric features appear in
the $K$-band spectrum of this Class~I object. However, in observations of
second overtone CO emission, Biscaya et al.\ (1999) detect absorption
in the $H$-band Brackett series of hydrogen, presumably arising from 
the stellar photosphere of the central star.
They conclude that the spectral type of WL16 is between B8 and A7
($T_{\rm eff}=8000$-12000~K). Subsequent modeling of the CO emission 
further constrains the temperature of the star to be 9000 to 12000~K.

GY20/DoAr24E is a binary system with a separation of $2\farcs06$ and
component magnitudes of $K=6.9$ and 7.9 (Simon et al.\ 1995). The
curvature in the slope of the Class~I secondary is due to its proximity to the 
primary and the resulting difficulties in extracting the spectrum. 
Since the companion is much redder than the primary, we assume the BKLT
$J$ photometry applies to only GY20A when deriving the luminosity. 

For GY29, both the CO bands and the shape of the continuum produced by steam 
absorption match well with the spectrum of V410~X-ray~6 (M5.5) artificially 
reddened by $A_V=25$. 

In addition to absorption in Br$\gamma$, we clearly detect weak features
of Na, Ca, and CO in the B star GY70/S1. The source of these metal lines
is likely the fainter, unresolved secondary ($\Delta K=1.8$, $0\farcs02$, 
Simon et al.\ 1995).

GY240/WL20 is a triple system (BKLT, references therein) appearing as one
source in the data of BKLT and Wilking \& Lada (1989) and 
two sources separated by 2-$3\arcsec$ in the images of CRBR and Barsony
et al.\ (1989). We refer to the west and east components as GY240A and B.
Since GY240B has no $J$ photometry and the $H-K$ colors of the two sources
are similar, we derive the extinction from BKLT $J-H$ for the
composite system and assume it applies to each component. To estimate 
a $J$ magnitude for the individual sources, we assume a difference of 
$\sim1.2$~mag, which is measured by CRBR at $H$ and $K$.

GY250/SR12 is a binary system with a separation of $0\farcs3$ and component
magnitudes of $K=9.3$ and 9.4 (Simon et al.\ 1995). We assume the components
have the same spectral types and divide the luminosity in half. Only one
source is added to the IMF, for reasons discussed in \S~\ref{sec:oph.complete}.

GY372/VSSG14 is an unresolved binary system ($0\farcs1$, Simon et al.\ 1995)
where the secondary
is 0.8~mag fainter at $K$. As with GY70/S1, Br$\gamma$ is strong while we
also detect absorption in metal lines
and CO that may arise in the secondary. Due to possible line and
continuum veiling (from either the secondary or IR excess emission), 
the classification of the primary is uncertain ($<$F8).

\newpage

\clearpage

\clearpage

\clearpage

\clearpage

\clearpage

\clearpage

\clearpage

\clearpage

\clearpage

\clearpage

\clearpage

\clearpage

\clearpage

\clearpage

\begin{deluxetable}{llllll}
\tablewidth{0pt}
\tablenum{2}
\tablecaption{SED Classifications and Spectroscopic Data}
\tablehead{\colhead{ID} & \multicolumn{2}{c}{$a$(2-10~$\mu$m)\tablenotemark{a}}
& \colhead{Class\tablenotemark{b}} & 
\colhead{$r_K$, Br$\gamma$, $a$(2-20~$\mu$m)\tablenotemark{c}} \\
\cline{2-3}
\colhead{} & \colhead{observed} & \colhead{dereddened} & \colhead{} & \colhead{}
}
\startdata
GY6 & 1.2 & \nodata & I & $>2$, 2.0$\pm$0.2~\AA, $a20=1.3$ \nl
GY111 & 1.2 & \nodata & I & 1-3, 1.8$\pm$0.4~\AA, $a20=0.9$ \nl
GY214 & 1.2 & \nodata & I & $>2$, 3.0$\pm$0.1~\AA, $a20=0.9$ \nl
GY254 & 1.0 & \nodata & I & $>1$, 6.2$\pm$0.4~\AA, $a20=0.6$ \nl
GY265 & 0.7 & \nodata & I & $>1$, 1.5$\pm$0.7~\AA, $a20=0.9$ \nl
GY269 & 1.4 & \nodata & I & 1, $a20=1.6$ \nl
GY274 & 1.4 & \nodata & I & $>1$, 2.0$\pm$0.6~\AA, $a20=1.1$ \nl
GY378 & 2.7 & \nodata & I & $>2$, 4.3$\pm$0.4~\AA, $a20=2$ \nl
GY205 & 0.3 & \nodata & flat & $>0.5$, 5.9$\pm$0.4~\AA, $a20<0.4$ \nl
GY224 & 0.2 & \nodata & flat & $>0.5$, 8.5$\pm$1.0~\AA, $a20=0.1$ \nl
GY227 & 0.2 & \nodata & flat & $>0.5$ \nl
GSS26 & 0.3 & $-0.4$ & flat/II & 4, 0.75, $a20=-0.1$ \nl
      &     &        &         & 2.9$\pm$0.2, 1.7$\pm$0.3~\AA \nl
GY21 & 0.0 & $-0.5$ & flat/II & 1\tablenotemark{d}, 2.0$\pm$0.7~\AA \nl
GY51 & 0.3 & $\sim-0.4$ & flat/II & 2, 0.5, 5.0$\pm$0.5, 1.5$\pm$0.5~\AA \nl
GY279 & $-0.2$, 0.3 & $-1.0$, $-0.5$ & flat/II & 0.75, 1\tablenotemark{d}, $a20=0$\tablenotemark{e} \nl
GY315 & 0.2 & $\sim-0.9$ & flat/II & 1-3, 2.0$\pm$0.6~\AA, $a20=0.1$ \nl
GY81 & \nodata & \nodata & flat/II? & 1-2, 3.0$\pm$0.5~\AA\ \nl
EL24 & $-0.7$ & $-0.9$ & II & 1-3, 2.7$\pm$0.2~\AA, $a20=-0.6$ \nl
GSS28 & $-0.8$ & $-0.9$ & II & 0.5, 1.9$\pm$0.15~\AA\ \nl
GSS29 & $-0.8$ & $-1.2$ & II & 0, $a20<-0.8$ \nl
GY23 & $-0.5$ & $-0.9$ & II & 0.75, $a20=-0.6$ \nl
GY110 & $-1.1$ & $-1.5$ & II & 0.25, 0.6$\pm$0.2\AA\ \nl
GY116 & $-0.7$ & $-1.3$ & II & 0.75, 2.1$\pm$0.15~\AA\ \nl
GY129 & $-1.1$ & $-1.4$ & II & $>1$, 13.7$\pm$0.5~\AA\ \nl
GY167 & $-0.6$ & $-0.7$ & II & 1-3, 4.7$\pm$0.15~\AA, $a20=-0.3$ \nl
GY168 & $-0.6$ & $-0.8$ & II & 2, 1.3$\pm$0.2,$a20=-0.5$ \nl
GY211 & $-1.0$ & $-1.4$ & II & 0.75, $a20<-0.3$ \nl
GY244 & 0.3 & $-1$?\tablenotemark{g} & II? & 0, $a20<0.5$ \nl
GY247 & $-0.9$ & $-1.5$ & II & 0.25, $a20<-0.9$ \nl
GY267 & $-1.1$ & $-1.5$ & II & 0.25 \nl
GY273 & $-0.3$ & $-1.0$ & II & 0, $a20<-0.1$ \nl
GY292 & $-0.8$ & $-1.1$ & II & 1, 1.0$\pm$0.4, 3.0$\pm$1~\AA\ \nl
GY308 & $-0.7$ & $-1.1$ & II & 0.75, 0.8$\pm$0.2~\AA\ \nl
GY314 & $-0.8$ & $-1.1$ & II & 0.5, 1.6$\pm$0.3, $<$0.5~\AA\ \nl
GY319 & $-1.3$ & $-1.3$ & II & 0.75, $a20=-1.1$ \nl
GY400 & $-1.1$ & $-1.1$ & II & 0.5, 2.5$\pm$0.3~\AA\ \nl
SR4 & $-1.0$ & $-1.1$ & II & 1.5, 2, 1.5, $a20=-0.7$ \nl
    &        &        &    & 6.7$\pm$0.3, 4.5$\pm$0.5, 5.5$\pm$0.5~\AA \nl   
IRS2 & $-1.4$ & $-1.6$ & II & 0 \nl
WSB60 & $-0.4$ & $-0.5$ & II & 0.75 \nl
GSS20 & $<-1.9$ & $<-2.0$ & III & 0 \nl
GSS23 & $-2.1$ & $-2.3$ & III & 0, $a20<-1.6$ \nl
GY5 & $\sim-1.5$\tablenotemark{h} & $\sim-1.9$\tablenotemark{h} & III & $<1$ \nl
GY12 & $<-2.0$ & $<-2.6$ & III & 0 \nl
GY17 & $-1.9$ & $-1.9$ & III & 0, 0.9$\pm$0.2~\AA\ \nl
GY29 & \nodata & \nodata & III?\tablenotemark{f} & 0 \nl
GY37 & \nodata & \nodata & III?\tablenotemark{f} & ? \nl
GY59 & \nodata & \nodata & III?\tablenotemark{f} & ? \nl
GY84 & \nodata & \nodata & III?\tablenotemark{f} & 0 \nl
GY135 & $<-2.1$ & $<-2.6$ & III & 0 \nl
GY153 & $-0.9$ & $-1.7$ & III & 0.25, 1.7$\pm$0.4~\AA\ \nl
GY156 & $<-2.2$ & $<-2.9$ & III & 0 \nl
GY172 & $<0.6$ & $<0.0$ & III? & 0, $a20<1.1$ \nl
GY250 & $<-2.10$ & $<-2.12$ & III & 0 \nl
GY253 & $<-1.2$ & $<-2.1$ & III & 0 \nl
GY262 & $-1.5$ & $-2.3$ & III & 1, 2.2$\pm$0.3~\AA\ \nl
GY284 & \nodata & \nodata & III?\tablenotemark{f} & 0 \nl
GY295 & $<-2.2$ & $<-2.4$ & III & 0 \nl
GY306 & $<-1.35$ & $<-1.73$ & III & 0 \nl
GY310 & \nodata & \nodata & III?\tablenotemark{f} & \nodata \nl
GY326 & \nodata & \nodata & III?\tablenotemark{f} & \nodata \nl
GY410 & $<-1.7$ & $<-2.0$ & III & 0 \nl
GY93 & $-1.7$ & $-1.8$ & III & 0 \nl
LFAM8 & \nodata & \nodata & III?\tablenotemark{f} & 0 \nl
ROXs39 & $-2.6$ & $-2.6$ & III & 0 \nl
SR20 & $-1.6$ & $-1.8$ & III & 1 \nl
VSSG11 & $<-1.12$ & $<-1.62$ & III & 0 \nl
VSSG22 & $<-1.6$ & $<-2.2$ & III & 0 \nl
SR22 & $-1.6$ & $-1.7$ & III & 0.25, 1.9$\pm$0.4~\AA\ \nl
WSB38 & $-1.6$ & $-1.9$ & III & 0.75, 0.6$\pm$0.15~\AA\ \nl
162559-242124 & \nodata & \nodata & III?\tablenotemark{f} & 0 \nl
162615-241924 & \nodata & \nodata & III?\tablenotemark{f} & 0 \nl
162618-242416 & $\sim-0.2$\tablenotemark{i} & $\sim-2.0$\tablenotemark{i} & III? & CRCTL \nl
\enddata
\tablenotetext{a}{Data taken from Elias (1978), Lada \& Wilking (1984),
Young, Lada, \& Wilking (1986), Wilking, Lada, \& Young (1989), and Greene 
et al.\ (1994).}
\tablenotetext{b}{SED classification (Greene et al.\ 1994) based on 
$a$(2-10~\micron) and the other data given (see \S~\ref{sec:oph.revised}).}
\tablenotetext{c}{Continuum veiling and $W_{\lambda}({\rm Br}\gamma)$ measured 
in the $K$-band spectra and the
spectral index from 2-20~\micron\ when available.}
\tablenotetext{d}{Greene \& Lada (1997).}
\tablenotetext{e}{Measurement of 20~\micron\ photometry was at the same epoch 
as the 10~\micron\ data used in $a10=0.3$.}
\tablenotetext{f}{Tentative SED classification based on $JHK$ colors and 
veiling in the $K$-band spectra.}
\tablenotetext{g}{Dereddening is based on the assumption that $H-K$ is not
affected significantly by excess emission, as suggested by $r_K\sim0$ 
measured in the spectrum.}
\tablenotetext{h}{$a$(2-6.0~$\mu$m).}
\tablenotetext{i}{$a$(2-4.5~$\mu$m).}
\end{deluxetable}


\begin{references}
\reference{} Allard, F., \& Hauschildt, P. H. 1995, \apj, 445, 433
\reference{} Baraffe, I., Chabrier, G., Allard, F., \& Hauschildt, P. H. 1997,
\aap, 327, 1054
\reference{} Barsony, M., Burton, M. G., Russel, A. P. G., Carlstrom, J. E., 
\& Garden, R. 1989, \apj, 346, L93 
\reference{} Barsony, M., Kenyon, S. J., Lada, E. A., \& Teuben, P. J. 1997,
\apjs, 112, 109 (BKLT)
\reference{} Biscaya, A. M., Calvet, N., Rieke, G. H., \& Luhman, K. L. 
in preparation
\reference{} Bouvier, J., \& Appenzeller, I. 1992, \aaps, 92, 481 
\reference{} Bouvier, J., Stauffer, J. R., Mart{\'\i}n, E. L., Barrado y 
Navascu\'{e}s, D., Wallace, B., \& Bejar, V. J. S. 1998, \aap, 336, 490
\reference{} Brice\~{n}o, C., Hartmann, L., Stauffer, J., \& Mart{\'\i}n, E. L.,
1998, \aj, 115, 2074
\reference{} Burrows, A., Marley, M., Hubbard, W. B., Lunine, J. I., Guillot,
T., Saumon, D., Freedman, R., Sudarsky, D., \& Sharp, C. 1997, \apj, 491, 856
\reference{} Calvet, N., Hartmann, L., \& Strom, S. E. 1997, \apj, 481, 912 
\reference{} Carr, J. S., Tokunaga, A. T., Najita, J., Shu, F. H., \& Glassgold,
A. E. 1993, \apj, 411, 37
\reference{} Casali, M. M., \& Eiroa, C. 1996, \aap, 306, 427
\reference{} Casali, M. M., \& Matthews, H. E. 1992, \mnras, 258, 399
\reference{} Casanova, S., Montmerle, T., Feigelson, E. D., \& Andr\'{e}, P.
1995, \apj, 439, 752
\reference{} Cohen, M., \& Kuhi, L. V. 1979, \apjs, 41, 743 
\reference{} Comer\'{o}n, F., Rieke, G. H., Burrows, A., \& Rieke, M. J. 1993,
\apj, 416, 185 (CRBR)
\reference{} Comer\'{o}n, F., Rieke, G. H., Claes, P., Torra, J., \& Laureijs,
R.  J., 1998, \aap, 335, 552 (CRCTL)
\reference{} Comer\'{o}n, F., Rieke, G. H., \& Neuh\"{a}user, R. 1999, \aap, 
343, 477 
\reference{} D'Antona, F., \& Mazzitelli, I. 1994, \apjs, 90, 467 (DM94)
\reference{} D'Antona, F., \& Mazzitelli, I. 1997, in ``Cool stars in
Clusters and Associations'', eds. R. Pallavicini \& G. Micela, Mem.S.A.It., 
68, n.4 (DM97)
\reference{} Dolidze, M. V., \& Arakeylan, M. A. 1959, Soviet Astron, 3, 434
\reference{} Elias, J. H. 1978, \apj, 224, 453 
\reference{} Elias, J. H., Frogel, J. A., Matthews, K., \& Neugebauer, G. 1982,
\aj, 87, 1029
\reference{} Grasdalen, G. L., Strom, K. M., \& Strom, K. E. 1973, \apj, 184, 
L53
\reference{} Greene, T. P., \& Lada, C. J., 1996, \aj, 112, 2184
\reference{} Greene, T. P., \& Lada, C. J., 1997, \aj, 114, 2157
\reference{} Greene, T. P., \& Meyer, M. R. 1995, \apj, 450, 233
\apj, 492, 323
\reference{} Greene, T. P., Wilking, B. A., Andr\'{e}, P., Young, E. T., \& 
Lada, C. J. 1994, \apj, 434, 614 
\reference{} Greene, T. P., \& Young, E. T. 1992, \apj, 395, 516
\reference{} Hartmann, L., \& Kenyon, S. J. 1996, \araa, 34, 207
\reference{} Herbig, G. H. 1998, \apj, 497, 736
\reference{} Hillenbrand, L. A. 1997, \aj, 113, 1733
\reference{} Humphreys, R. M., Jones, T. J., \& Sitko, M. L. 1984, \aj, 89, 1155
\reference{} Kenyon, S. J., Brown, D. I., Tout, C. A., \& Berlind, P. 1998a, 
\aj, 115, 2491
\reference{} Kenyon, S. J., \& Hartmann, L. 1995, \apjs, 101, 117
\reference{} Kenyon, S. J., Lada, E. A., \& Barsony, M. 1998b, \aj, 115, 252
\reference{} Kirkpatrick, J. D., Henry, T. J., \& Irwin, M. J. 1997, \aj, 113, 
1421
\reference{} Kirkpatrick, J. D., et al. 1999, \apj, in press
\reference{} Kleinmann, S. G., \& Hall, D. N. B. 1986, \apjs, 62, 501
\reference{} Lada, C. J. 1987, in Star Forming Regions, ed. M. Peimbert and 
J. Jugaku (Dordrecht: Reidel), p. 1
\reference{} Lada, C. J.,\& F. C. Adams, 1992, \apj, 393, 278
\reference{} Lada, C. J., \& Wilking, B. A. 1984, \apj, 287, 610
\reference{} Leggett, S. K. 1992, \apjs, 82, 351
\reference{} Leous, J. A., Feigelson, E. D., Andr\'{e}, P., \& Montmerle, T.
1991, \apj, 379, 683
\reference{} Luhman, K. L. 1998, in ASP Conf. Ser. 134, Brown Dwarfs and
Extrasolar Planets Proceedings, ed. R. Rebolo, E. L. Mart{\'\i}n, M. R.
Zapatero-Osorio (San Francisco: ASP), 532
\reference{} Luhman, K. L. 1999, \apj, submitted
\reference{} Luhman, K. L., Brice\~{n}o, C., Rieke, G. H., \& Hartmann, L. W.
1998a, \apj, 493, 909
\reference{} Luhman, K. L., Liebert, J., \& Rieke, G. H. 1997, \apj, 489, L165 
\reference{} Luhman, K. L., \& Rieke, G. H. 1998, \apj, 497, 354 (LR)
\reference{} Luhman, K. L., Rieke, G. H., Lada, C. J., \& Lada, E. A. 1998b, 
\apj, 508, 347 (LRLL)
\reference{} Mart{\'\i}n, E. L., Montmerle, T., Gregorio-Hetem, J., \& Casanova,
S. 1998, \mnras, 300, 733
\reference{} Mart{\'\i}n, E. L., Rebolo, R., \& Zapatero Osorio, M. R. 1996,
\apj, 469, 706
\reference{} Meyer, M. R. 1995, Ph.D. thesis, Univ. Massachusetts
\reference{} Meyer, M. R., Calvet, N., \& Hillenbrand, L. A. 1997, \aj, 114, 288
\reference{} Miller, G. E., \& Scalo, J. M. 1979, \apjs, 41, 513
\reference{} Montmerle, T., Koch-Miramonde, L., Falgarone, E., \& Grindlay, 
J. E., 1983, \apj, 269, 182
\reference{} Motte, F., Andr\'{e}, P., \& Neri, R. 1998, \aap, 336, 150
\reference{} Muzerolle, J., Calvet, N., \& Hartmann, L. 1998, \apj, 492, 743
\reference{} Myers, P. C., Fuller, G. A., Mathieu, R. D., Beichman, C. A., 
Benson, P. J., Schild, R. E., \& Emerson, J. P. 1987, \apj, 319, 340
\reference{} Najita, J., Carr, J. S., Glassgold, A. E., Shu, F. H., \& 
Tokunaga, A. T. 1996, \apj, 462, 919
\reference{} Reid, I. N., et al. 1999, \apj, submitted
\reference{} Rieke, G. H., Ashok, N. M., \& Boyle, R. P. 1989, \apj, 339, L71
\reference{} Rieke, G. H., \& Lebofsky, M. J. 1985, \apj, 288, 618
\reference{} Rieke, G. H., \& Rieke, M. J. 1990, \apj, 362, L21
\reference{} Salpeter, E. E. 1955, \apj, 121, 161
\reference{} Scalo, J. 1986, Fund. Cosmic Phys., 11, 1
\reference{} Scalo, J. 1998, ASP Conf. Ser. 142, ed. G. Gilmore, I. Parry \& S. 
Ryan, 201
\reference{} Shu, F. H., Adams, F. C., \& Lizano, S. 1987, \araa, 25, 23
\reference{} Simon, M., et al. 1995, \apj, 443, 625
\reference{} Stahler, S. W. 1983, \apj, 274, 822
\reference{} Strom, K. M., Kepner, J., \& Strom, S. E. 1995, \apj, 438, 813
\reference{} Strom, K. M., \& Strom, S. E. 1994, \apj, 424, 237 
\reference{} Struve, O., \& Rudkj{\o}bing, M. 1949, \apj, 109, 92
\reference{} Testi, L., \& Sargent, A. I. 1998, \apj, 508, L91
\reference{} Tokunaga, A. T., \& Kobayashi, N. 1999, \aj, 117, 1010
\reference{} Vrba, F. J., Strom, S. E., \& Strom, K. M. 1976, \aj, 81, 958
\reference{} Vrba, F. J., Strom, K. M., Strom, S. E., \& Grasdalen, G. L. 1975,
\apj, 197, 77
\reference{} Wainscoat, R. J., Cohen, M. Volk, K., Walker, H. J., Schwartz, D.
E. 1992, \apjs, 83, 111
\reference{} Wallace, L., \& Hinkle, K. 1996, \apjs, 107, 312
\reference{} Whittet, D. C. B. 1974, \mnras, 168, 371
\reference{} Wilking, B. A., Greene, T. P., \& Meyer, M. R. 1998, \aj, 117, 469
(WGM)
\reference{} Wilking, B. A., \& Lada, C. J. 1983, \apj, 274, 698
\reference{} Wilking, B. A., Lada, C. J., \& Young, E. T. 1989, \apj, 340, 823
\reference{} Wilking, B. A., Schwartz, R. D., \& Blackwell, J. H. 1987, \aj, 
94, 106
\reference{} Williams, D. M., Comer\'{o}n, F., Rieke, G. H., \& Rieke, M. J. 
1995, \apj, 454, 144
\reference{} Williams, D., Thompson, C. L., Rieke, G. H, \& Montgomery, E. F.
1993, ProcSPIE, 1308, 482
\reference{} Young, E. T., Lada, C. J., \& Wilking, B. A. 1986, \apj, 304, L45
\end{references}
\end{document}


\pagestyle{empty}
\begin{deluxetable}{lllllllllllllll}
\scriptsize
\tablewidth{0pt}
\tablecaption{Data for Spectroscopic Sample in $\rho$ Ophiuchi}
\tablehead{\colhead{BKLT\tablenotemark{a}} & \colhead{Other ID\tablenotemark{b}} &
\colhead{$\alpha$(1950)} & \colhead{$\delta$(1950)} &
\colhead{previous\tablenotemark{c}} & \colhead{IR} &
\colhead{adopt\tablenotemark{d}} & \colhead{$T_{\rm eff}$} &
\colhead{$A_{J}$} & \colhead{$L_{\rm bol}$} &
\colhead{$J-H$} & \colhead{$H-K$} & \colhead{$K$} &
\colhead{$r_{K}$\tablenotemark{e}} & \colhead{Br$\gamma$}
\\
\colhead{} & \colhead{} &
\colhead{min sec} & \colhead{$\arcmin$  $\arcsec$} &
\colhead{} & \colhead{} &
\colhead{} & \colhead{(K)} &
\colhead{} & \colhead{($L_{\odot}$)} &
\colhead{} & \colhead{} & \colhead{} &
\colhead{} & \colhead{(\AA)}
}
\startdata
\nodata & SR22 & 22\ 22.7 & 22\ 55.0 & K7-M0(GM),M0(CK),M3.5(M) & K6-M2 & M0 & 3850 & 0.6 & 0.76 & 0.93 & 0.45 & 8.70 & 0.25 & 1.9$\pm$0.4 \nl
162536-241544 & IRS2 & 22\ 35.59 & 08\ 57.6 & K7(GM) & K3-K4 & K3.5 & 4660 & 1.9 & 2.1 & 1.39 & 0.70 & 8.36 & 0 & 0 \nl
162556-242050 & SR4,EL13,YLW25,IRS12 & 22\ 54.82 & 14\ 04.5 & K6(BA),K7(CK) & K5-M2 & K8 & 3955 & 0.5 & 1.9 & 1.06 & 0.72 & 7.25 & 1.5,2,1.5 & 6.7$\pm$0.3,4.5$\pm$0.5  \nl
 & ROXs6,WSB25,DoAr20 &  &  &  &  &  &  &  &  &  &  &  &  & 5.5$\pm$0.5 \nl
162557-243032 & GSS20,IRS13,ROXs7 & 22\ 56.11 & 23\ 46.6 & K7(GM),K7(BA) & K5-K7 & K6 & 4205 & 0.9 & 1.1 & 1.12 & 0.63 & 8.37 & 0 & 0 \nl
162558-242203 & \nodata & 22\ 57.23 & 15\ 17.7 & \nodata & K0-K2III: & b? & \nodata & \nodata & \nodata & 2.57 & 1.46 & 11.41 & ? & 0  \nl
162559-242124 & \nodata & 22\ 58.32 & 14\ 38.3 & \nodata & M4 & M4 & 3180 & 0.0 & 0.21 & 0.65 & 0.45 & 9.52 & 0 & 0  \nl
162601-242521 & \nodata & 23\ 00.01 & 18\ 36.2 & \nodata & K-M: & \nodata & \nodata & \nodata & \nodata & 1.90 & 1.23 & 11.10 & ? & 0  \nl
162603-242337 & GSS23,DoAr21,EL14 & 23\ 01.65 & 16\ 52.4 & K0(BA),K0(M) & K0-K2 & K1 & 5080 & 1.6 & 18 & 1.26 & 0.59 & 6.16 & 0 & 0 \nl
 & ROXs8,VSS93 &  &  &  &  &  &  &  &  &  &  &  &  &  \nl
162605-242451(c) & \nodata & 23\ 03.64 & 18\ 05.9 & early(WGM) & \nodata & b & \nodata & \nodata & \nodata & >1.1 & 1.91 & 13.99 & \nodata & \nodata  \nl
162609-243411 & GSS25,VSS92,EL16 & 23\ 07.84 & 27\ 26.6 & A0(SR) & $<$F3 & A0 & 9520 & 2.3 & 160 & 0.86 & 0.49 & 6.42 & 0 & 0 \nl
 & YLW28,SR3 &  &  &  &  &  &  &  &  &  &  &  &  &  \nl
162610-242056 & GSS26 & 23\ 08.97 & 14\ 11.6 & \nodata & K5-M2 & K8 & 3955 & 6.0 & 1.3 & 3.34 & 2.25 & 9.38 & 4,0.75 & 2.9$\pm$0.2,1.7$\pm$0.3 \nl
162615-241924(c) & \nodata & 23\ 14.49 & 12\ 40.0 & \nodata & M0-M3 & M1.5 & 3595 & 4.6 & 0.80 & 2.51 & 1.46 & 9.98 & 0 & 0  \nl
162616-242225(c) & GSS29,EL18 & 23\ 15.46 & 15\ 40.7 & K7-M0(GM) & K5-K7 & K6 & 4205 & 3.0 & 3.1 & 1.88 & 1.04 & 8.19 & 0 & 0 \nl
162617-242023(c) & GSS28,DoAr24,EL19 & 23\ 15.71 & 13\ 39.3 & K7-M0(GM),K5(BA) & K6-M2 & M0 & 3850 & 0.8 & 1.2 & 1.05 & 0.58 & 8.09 & 0.5 & 1.9$\pm$0.15 \nl
 & ROXs10A,WSB27 &  &  &  &  &  &  &  &  &  &  &  &  &  \nl
162618-242818 & \nodata & 23\ 16.61 & 21\ 33.6 & early(WGM) & \nodata & b & \nodata & \nodata & \nodata & 2.57 & 1.43 & 12.93 & \nodata & \nodata  \nl
162618-242611(c) & \nodata & 23\ 17.54 & 19\ 27.4 & M7.5(WGM) & M5-M6\tablenotemark{f} & M5.5 & 2925 & 3.1 & 0.050 & 1.75 & 1.20 & 12.18 & \nodata & \nodata  \nl
162618-242416(c) & \nodata & 23\ 17.56 & 17\ 32.2 & M5(WGM) & K-M & M5 & 3010 & 5.0 & 0.097 & 2.46 & 1.95 & 11.94 & \nodata & \nodata  \nl
162621-242306(c) & GY6,GSS30,EL21 & 23\ 19.99 & 16\ 22.1 & \nodata & ? & \nodata & \nodata & \nodata & \nodata & 3.06 & 2.51 & 8.32 & $>$2 & 2.0$\pm$0.2,$<$0.5 \nl
162621-242602(c) & GY5 & 23\ 20.27 & 19\ 17.7 & M7(WGM) & M5-M6\tablenotemark{f} & M5.5 & 2925 & 1.4 & 0.11 & 1.13 & 0.66 & 10.91 & $<1$ & \nodata  \nl
162622-242354(c) & GY10 & 23\ 20.78 & 17\ 10.2 & M8.5(WGM) & M6-M7\tablenotemark{f} & M6.5 & 2780 & 4.0 & 0.067 & 2.16 & 1.34 & 12.25 & \nodata & \nodata  \nl
162622-242409(c) & GY11 & 23\ 20.89 & 17\ 25.1 & M6.5(WGM) & \nodata & M6.5 & 2780 & 1.4 & 0.0028 & 1.15 & 1.22 & 14.15 & \nodata & \nodata  \nl
162622-242254(c) & GY12 & 23\ 20.99 & 16\ 10.8 & M2(GM) & K6-M1 & K8 & 3955 & 4.6 & 0.68 & 2.75 & 1.87 & 9.60 & 0 & 0 \nl
162623-242101(c)\tablenotemark{g} & GY20A,DoAr24E,GSS31 & 23\ 22.01 & 14\ 17.6 & K0(BA) & K0-K2 & K1 & 5080 & 1.8 & 10 & 1.52 & 0.93 & 6.44 & 0 & 0.7$\pm$0.2 \nl
 & EL22,ROXs10B,WSB30 &  &  &  &  &  &  &  &  &  &  &  &  &  \nl
162623-242101(c)\tablenotemark{g} & GY20B,DoAr24EB & 23\ 22.01 & 14\ 17.6 & \nodata & ? & \nodata & \nodata & \nodata & \nodata & 1.52 & 0.93 & 6.44 & $>$2 & 1.5$\pm$0.4 \nl
162623-244311 & GY17,DoAr25 & 23\ 22.04 & 36\ 27.7 & K4(GM),K7(M) & K5-K7 & K6 & 4205 & 0.0 & 1.0 & 0.93 & 0.79 & 7.57 & 0 & 0.9$\pm$0.2 \nl
162623-242441(c) & GY21,LFAM3 & 23\ 22.14 & 17\ 57.4 & K-M(GM),M0(GL) & K-M: & M0 & 3850 & 3.8 & 0.36 & 2.45 & 1.69 & 9.94 & 1(GL) & 2.0$\pm$0.7  \nl
162624-242449(c) & GY23,S2,GSS32 & 23\ 22.64 & 18\ 05.6 & K(GM) & K5-M2 & K8 & 3955 & 3.1 & 3.9 & 2.15 & 1.48 & 7.20 & 0.75 & 0  \nl
 & EL23 &  &  &  &  &  &  &  &  &  &  &  &  &  \nl
162624-241616 & EL24,YLW32 & 23\ 22.85 & 09\ 32.0 & \nodata & G5-K7 & \nodata & \nodata & 2.3 & \nodata & 1.95 & 1.46 & 6.77 & 1-3 & 2.7$\pm$0.2 \nl
162625-242325(c) & GY31 & 23\ 23.86 & 16\ 41.7 & M5.5(WGM) & \nodata & M5.5 & 2925 & 16 & 1.0 & \nodata & 3.98 & 13.09 & \nodata & \nodata  \nl
162625-242446(c) & GY29 & 23\ 23.87 & 18\ 02.7 & \nodata & M4-M6 & M5 & 3010 & 6.3 & 0.47 & 2.95 & 2.11 & 10.86 & 0 & 0  \nl
162627-242643(c) & GY37,VSSG29 & 23\ 26.38 & 19\ 59.5 & M6(WGM) & M6\tablenotemark{f} & M6 & 2840 & 1.8 & 0.037 & 1.31 & 0.95 & 11.99 & \nodata & \nodata  \nl
162629-241908 & LFAM8 & 23\ 28.31 & 12\ 24.3 & \nodata & M0-M3 & M1.5 & 3595 & 6.6 & 0.46 & 3.37 & 2.05 & 11.19 & 0 & 0  \nl
162630-243841 & GY45 & 23\ 28.45 & 31\ 57.4 & \nodata & G8.5III & b & \nodata & \nodata & \nodata & 2.40 & 1.36 & 8.31 & \nodata & 0 \nl
162630-242258(c) & GY51,VSSG27 & 23\ 29.12 & 16\ 15.3 & \nodata & K3-M1 & \nodata & \nodata & \nodata & \nodata & 3.23 & 2.74 & 10.72 & 2,0.5 & 5.0$\pm$0.5,1.5$\pm$0.5  \nl
162631-242531(c) & GY59 & 23\ 29.95 & 18\ 48.0 & M6(WGM) & M4-M6 & M5 & 3010 & 3.4 & 0.099 & 1.86 & 1.21 & 11.68 & \nodata & \nodata  \nl
162632-242636(c) & GY64 & 23\ 31.11 & 19\ 53.4 & M8(WGM) & \nodata & M7 & 2720 & 3.2 & 0.014 & 1.87 & 1.43 & 13.33 & \nodata & \nodata  \nl
162633-243625 & GY65 & 23\ 31.34 & 29\ 42.4 & \nodata & M2-3III & b & \nodata & \nodata & \nodata & 2.58 & 1.49 & 8.91 & \nodata & 0 \nl
162634-242330(c) & GY70,S1,GSS35,VSS26 & 23\ 32.78 & 16\ 47.1 & B2-B5(BA,GSS,CK) & $<$F8 & B3 & 18700 & \nodata & \nodata & 1.56 & 0.96 & 6.32 & 0 & 0  \nl
 & EL25,ROXs14,YLW36 &  &  &  &  &  &  &  &  &  &  &  &  &  \nl
162637-242302 & GY81,VSSG4 & 23\ 36.42 & 16\ 19.4 & \nodata & K5-M2 & K8 & 3955 & 4.6 & 0.20 & 2.79 & 1.91 & 10.92 & 1-2 & 3.0$\pm$0.5  \nl
162638-242324 & GY84 & 23\ 37.43 & 16\ 41.6 & M6(WGM) & M0-M3 & M1.5 & 3595 & 4.6 & 0.24 & 2.40 & 1.26 & 11.63 & 0 & 0  \nl
162641-244015 & GY93,WSB37 & 23\ 39.72 & 33\ 33.0 & K-M(GM) & M4 & M4 & 3180 & 0.6 & 0.31 & 0.98 & 0.55 & 9.20 & 0 & 0 \nl
162642-242224 & GY107 & 23\ 41.32 & 15\ 42.0 & M3?(WGM) & \nodata & b & \nodata & \nodata & \nodata & 2.02 & 0.97 & 11.40 & \nodata & \nodata  \nl
162642-242031 & GY110,VSSG2,EL26,GSS37 & 23\ 41.48 & 13\ 49.0 & M0(GM),K6(M) & K6-M2 & M0 & 3850 & 2.9 & 3.5 & 1.78 & 0.93 & 8.00 & 0.25 & 0.6$\pm$0.2 \nl
162643-241635 & VSSG11,IRS19 & 23\ 42.42 & 09\ 52.9 & M3(GM) & K6-M2 & M0 & 3850 & 4.0 & 1.1 & 2.27 & 1.27 & 9.58 & 0 & 0 \nl
162644-243447 & GY111,WL12,YLW2 & 23\ 42.66 & 28\ 04.7 & \nodata & G5-K7 & \nodata & \nodata & \nodata & \nodata & 3.79 & 2.89 & 10.18 & 1-3 & 1.8$\pm$0.4  \nl
 & VSSG30 &  &  &  &  &  &  &  &  &  &  &  &  &  \nl
162645-242309 & GY116,GSS39,EL27 & 23\ 43.62 & 16\ 26.9 & M(GM) & K5-M2 & K8 & 3955 & 4.7 & 2.1 & 2.64 & 1.62 & 8.88 & 0.75 & 2.1$\pm$0.15 \nl
 & VSSG28 &  &  &  &  &  &  &  &  &  &  &  &  &  \nl
162646-241203 & WSB38,VSS27,ROXs16 & 23\ 45.16 & 05\ 20.9 & K(GM),G9:(BA) & K5-M2 & K8 & 3955 & 1.9 & 3.4 & 1.47 & 0.82 & 7.51 & 0.75 & 0.6$\pm$0.15 \nl
162648-242840(c) & GY128,WL2 & 23\ 47.05 & 21\ 57.9 & K-M(GM) & K-M: & \nodata & \nodata & \nodata & \nodata & $>$2.95 & 3.06 & 10.99 & ? & 0  \nl
162649-243823 & GY129,WL18 & 23\ 47.47 & 31\ 41.3 & \nodata & ? & \nodata & \nodata & 2.6 & \nodata & 1.99 & 1.39 & 9.82 & $>$1 & 13.7$\pm$0.5 \nl
162649-242005 & GY135,VSSG3,LFAM18 & 23\ 47.86 & 13\ 23.0 & K7-M0(GM) & K5-K7 & K6 & 4205 & 4.4 & 4.4 & 2.33 & 1.20 & 8.62 & 0 & 0 \nl
162651-243242(c) & GY141 & 23\ 49.82 & 26\ 00.4 & M8.5(LLR) & \nodata & M8.5 & 2500 & 0\tablenotemark{h} & 0.0026\tablenotemark{h} & 0.76 & 0.50 & 13.87 & \nodata & \nodata  \nl
162653-242229 & GY150,GSS40,VSSG6 & 23\ 52.47 & 15\ 47.6 & \nodata & K0-1III & b & \nodata & \nodata & \nodata & 2.86 & 1.67 & 9.85 & \nodata & 0  \nl
 & IRS21 &  &  &  &  &  &  &  &  &  &  &  &  &  \nl
162654-242622 & GY153,VSSG5 & 23\ 53.04 & 19\ 40.2 & M2(GM) & K6-M2 & M0 & 3850 & 5.8 & 1.2 & 3.02 & 1.82 & 9.88 & 0.25 & 1.7$\pm$0.4 \nl
162654-242231 & GY156,VSSG6,GSS40 & 23\ 53.58 & 15\ 49.4 & \nodata & K6-M1 & K8 & 3955 & 6.1 & 1.3 & 3.07 & 1.78 & 10.19 & 0 & 0 \nl
162656-242838(c) & GY163 & 23\ 55.47 & 21\ 56.8 & M2.5(WGM) & \nodata & M2.5 & 3430 & 9.0 & 0.26 & >2.06 & 2.28 & 12.66 & \nodata & \nodata  \nl
162658-244529 & GY168,SR24N,IRS24 & 23\ 56.58 & 38\ 48.1 & K2(GM),M0.5(CK),M0(M) & K5-M2 & K8 & 3955 & 2.0 & 2.3 & 1.70 & 1.18 & 7.42 & 2 & 1.3$\pm$0.2 \nl
 & EL28 &  &  &  &  &  &  &  &  &  &  &  &  &  \nl
162658-244534 & GY167,SR24S,IRS25 & 23\ 56.64 & 38\ 52.9 & K2(GM),K2:(CK),K7(M) & G5-K7 & K2 & 4900 & 1.2 & 2.4 & 1.50 & 1.17 & 7.08 & 1-3 & 4.7$\pm$0.15 \nl
 & EL28 &  &  &  &  &  &  &  &  &  &  &  &  &  \nl
162659-243556(c) & GY172,WL14 & 23\ 57.42 & 29\ 14.7 & \nodata & M4 & M4 & 3180 & 5.1 & 0.14 & 2.74 & 1.64 & 11.74 & 0 & 0  \nl
162702-243726(c) & GY182,WL16 & 24\ 00.64 & 30\ 45.0 & \nodata & B8-A7\tablenotemark{i} & \nodata & \nodata & \nodata & \nodata & 3.56 & 2.66 & 7.92 & ? & 7.2$\pm$0.2 \nl
162704-242830(c) & GY192,WL1,YLW6 & 24\ 02.61 & 21\ 49.3 & \nodata & $>$M3 & \nodata & \nodata & \nodata & \nodata & 3.15 & 2.26 & 10.81 & ? & 0  \nl
 & IRS29,VSSG31 &  &  &  &  &  &  &  &  &  &  &  &  &  \nl
162706-242837(c) & GY202 & 24\ 04.47 & 21\ 56.0 & M7(WGM) & M6-M7\tablenotemark{f} & M6.5 & 2780 & 3.9 & 0.022 & 2.09 & 1.70 & 12.97 & \nodata & \nodata  \nl
162706-243814(c) & GY205,WL17 & 24\ 05.06 & 31\ 33.3 & \nodata & ? & \nodata & \nodata & \nodata & \nodata & 3.33 & 3.29 & 10.28 & $>$0.5 & 5.9$\pm$0.4 \nl
162709-243408(c) & GY211,WL10 & 24\ 07.48 & 27\ 27.0 & \nodata & K5-M2 & K8 & 3955 & 3.6 & 1.4 & 2.22 & 1.39 & 8.85 & 0.75 & 1.4$\pm$0.2,$<$0.8  \nl
162709-243718(c) & GY214,EL29,WL15 & 24\ 07.71 & 30\ 37.4 & \nodata & ? & \nodata & \nodata & \nodata & \nodata & 5.20 & 4.47 & 7.54 & $>$2 & 3.0$\pm$0.1  \nl
 & YLW7 &  &  &  &  &  &  &  &  &  &  &  &  &  \nl
162710-241914 & SR21,VSSG23,VSS31 & 24\ 08.78 & 12\ 33.7 & G1(M) & F0-F8 & F4 & 6590 & 2.7 & 51 & 1.23 & 1.03 & 6.30 & $<$1 & 0 \nl
 & EL30,YLW8 &  &  &  &  &  &  &  &  &  &  &  &  &  \nl
162711-244046 & GY224 & 24\ 09.38 & 34\ 05.7 & \nodata & ? & \nodata & \nodata & \nodata & \nodata & $>$3.31 & 2.90 & 10.79 & $>$0.5 & 8.5$\pm$1.0 \nl
162711-243832(c) & GY227,WL19 & 24\ 10.00 & 31\ 51.2 & \nodata & ? & \nodata & \nodata & \nodata & \nodata & \nodata & $>$4.27 & 11.23 & $>$0.5 & 0 \nl
162711-242343 & GY228,IRS32 & 24\ 10.25 & 17\ 02.6 & \nodata & $<$F8 & \nodata & \nodata & \nodata & \nodata & 2.63 & 1.58 & 10.06 & ? & 0 \nl
162712-243449(c) & \nodata & 24\ 10.49 & 28\ 08.4 & \nodata & K-M: & \nodata & \nodata & \nodata & \nodata & 2.37 & 1.76 & 11.53 & ? & 0  \nl
162713-244133 & GY232 & 24\ 11.44 & 34\ 52.5 & \nodata & M0III & b & \nodata & \nodata & \nodata & $>$4.46 & 2.96 & 9.58 & \nodata & 0 \nl
162714-245132 & WSB45,ROXs20A & 24\ 12.34 & 44\ 51.7 & M5(BA) & M1-M4 & PMS?,dMe? & \nodata & \nodata & \nodata & 0.79 & 0.34 & 10.39 & 0 & 0  \nl
162715-245137 & WSB46,ROXs20B & 24\ 12.98 & 44\ 57.2 & M2(BA) & M0-M3 & PMS?,dMe? & \nodata & \nodata & \nodata & 0.90 & 0.46 & 9.51 & 0 & 0 \nl
162715-242640 & GY239,IRS34,YLW10B & 24\ 13.99 & 19\ 59.9 & K(GM) & K-M: & \nodata & \nodata & \nodata & \nodata & $>$3.53 & 2.81 & 10.66 & ? & 0  \nl
162715-243843(c)\tablenotemark{g} & GY240A,WL20 & 24\ 14.08 & 32\ 02.9 & K7-M0(GM) & K6-M2 & M0 & 3850 & 4.9 & 1.4 & 2.67 & 1.57 & 9.21 & 0.25 & 0 \nl
162715-243843(c)\tablenotemark{g} & GY240B,WL20 & 24\ 14.08 & 32\ 02.9 & K-M(GM) & K5-K7 & K6 & 4205 & 4.9 & 0.55 & 2.67 & 1.57 & 9.21 & 0 & 0  \nl
162717-242856(c) & GY244,YLW12A,IRS37 & 24\ 16.06 & 22\ 16.1 & \nodata & M4 & M4 & \nodata & $\sim11$\tablenotemark{j} & \nodata & $>$3.1 & 2.95 & 10.95 & 0 & 0  \nl
 & VSSG26 &  &  &  &  &  &  &  &  &  &  &  &  &  \nl
162718-242853(c) & GY246,WL5,VSSG26 & 24\ 16.65 & 22\ 12.6 & F7(GM) & $<$G0 & \nodata & \nodata & \nodata & \nodata & $>$2.68 & 4.04 & 10.28 & ? & 0 \nl
 & IRS38 &  &  &  &  &  &  &  &  &  &  &  &  &  \nl
162718-242906(c) & GY247,WL4,IRS39 & 24\ 16.95 & 22\ 25.6 & M(GM) & M0-M3 & M1.5 & 3595 & 5.4 & 1.6 & 3.00 & 1.93 & 9.13 & 0.25 & 0  \nl
 & VSSG26 &  &  &  &  &  &  &  &  &  &  &  &  &  \nl
162719-244139(c) & GY250,SR12,IRS40 & 24\ 17.67 & 34\ 59.5 & K4/M2.5(BA) & K6-M2 & M0 & 3850 & 0.2 & 0.66 & 0.73 & 0.28 & 8.41 & 0 & 0  \nl
 & ROXs21,YLW13A &  &  & M1(CK) &  &  &  &  &  &  &  &  &  &  \nl
162721-244142(c) & GY252,YLW13B,IRS42 & 24\ 19.63 & 35\ 02.5 & \nodata & $<$K0 & \nodata & \nodata & \nodata & \nodata & 3.90 & 2.90 & 8.41 & $>$1 & 0 \nl
162721-244335 & GY253 & 24\ 19.95 & 36\ 55.5 & \nodata & M2-M4 & M3 & 3350 & 8.2 & 0.83 & 4.03 & 2.53 & 10.78 & 0 & 0  \nl
162721-242953(c) & GY254,YLW14,WL6 & 24\ 20.26 & 23\ 12.9 & \nodata & ? & \nodata & \nodata & \nodata & \nodata & $>$2.61 & 4.35 & 10.04 & $>$1 & 6.2$\pm$0.4  \nl
162722-241759 & VSSG22 & 24\ 21.51 & 11\ 19.1 & K7-M0(GM) & K5-K7 & K6 & 4205 & 4.6 & 1.8 & 2.48 & 1.38 & 9.41 & 0 & 0 \nl
162726-243923(c) & GY262 & 24\ 24.67 & 32\ 44.0 & \nodata & K7-M3 & M1 & 3680 & 6.4 & 1.0 & 3.46 & 2.30 & 9.77 & 1 & 2.2$\pm$0.3  \nl
162726-244051(c) & GY265,YLW15A,IRS43 & 24\ 25.09 & 34\ 11.6 & \nodata & ? & \nodata & \nodata & \nodata & \nodata & $>$3.83 & 3.71 & 9.46 & $>$1 & 1.5$\pm$0.7,$<$0.5  \nl
162727-243116 & GY267,VSSG25,EL31 & 24\ 25.83 & 24\ 37.0 & K-M(GM) & K6-M2 & M0 & 3850 & 3.1 & 0.92 & 1.93 & 1.11 & 9.30 & 0.25 & 0 \nl
 & WL13 &  &  &  &  &  &  &  &  &  &  &  &  &  \nl
162728-243934 & GY269,YLW16A,IRS44 & 24\ 26.22 & 32\ 54.5 & \nodata & K5-M2 & K8 & \nodata & \nodata & \nodata & $>$3.91 & 3.44 & 9.65 & 1 & 0 \nl
162728-242721 & GY273,VSSG18,EL32,IRS45 & 24\ 26.92 & 20\ 42.1 & \nodata & K5-M0 & K6.5 & 4132 & 5.0 & 0.86 & 2.99 & 2.09 & 9.39 & 0 & 0 \nl
162729-243917 & GY274,YLW16B,IRS46 & 24\ 27.63 & 32\ 37.2 & \nodata & ? & \nodata & \nodata & \nodata & \nodata & $>$2.35 & 3.19 & 11.46 & $>$1 & 2.0$\pm$0.6 \nl
162730-242744 & GY279,VSSG17,EL33 & 24\ 28.65 & 21\ 04.6 & M(GM),$\geq$M2(GL) & K5-M2 & K8 & 3955 & 6.9 & 1.8 & 3.80 & 2.69 & 8.95 & 0.75,1(GL) & 0  \nl
 & IRS47 &  &  &  &  &  &  &  &  &  &  &  &  &  \nl
162730-242457 & GY284 & 24\ 29.37 & 18\ 17.4 & \nodata & M0-M3 & M1.5 & 3595 & 2.4 & 0.35 & 1.69 & 0.97 & 10.04 & 0 & 0  \nl
162733-244115(c) & GY292 & 24\ 31.29 & 34\ 36.2 & \nodata & K5-M2 & K8 & 3955 & 2.9 & 2.1 & 2.04 & 1.37 & 7.92 & 1 & 1.0$\pm$0.4,3.0$\pm$1 \nl
162735-243834 & GY295 & 24\ 33.50 & 31\ 55.0 & \nodata & M4 & M4 & 3180 & 1.3 & 0.36 & 1.13 & 0.51 & 9.64 & 0 & 0  \nl
162736-242833 & GY297 & 24\ 34.95 & 21\ 54.7 & \nodata & M0-M3 & f & \nodata & \nodata & \nodata & 0.68 & 0.25 & 11.04 & 0 & 0  \nl
162737-243035 & GY304,YLW46A,IRS48 & 24\ 35.63 & 23\ 56.0 & \nodata & $<$F3 & \nodata & \nodata & \nodata & \nodata & 1.88 & 1.23 & 7.42 & ? & 0  \nl
162738-243043 & GY306,IRS50 & 24\ 36.57 & 24\ 03.9 & \nodata & M4 & M4 & 3180 & 3.1 & 0.63 & 1.87 & 1.01 & 9.59 & 0 & 0  \nl
162738-243658 & GY308,IRS49 & 24\ 36.67 & 30\ 19.1 & \nodata & K5-M2 & K8 & 3955 & 2.9 & 1.7 & 1.96 & 1.24 & 8.31 & 0.75 & 0.8$\pm$0.2,$<$2  \nl
162738-243838 & GY310 & 24\ 36.90 & 31\ 59.1 & M8.5(WGM) & \nodata & M7 & 2720 & 1.7 & 0.079 & 1.29 & 0.83 & 11.08 & \nodata & \nodata  \nl
162739-243914 & GY314 & 24\ 37.68 & 32\ 35.6 & M(GM) & K6-M2 & M0 & 3850 & 2.0 & 1.4 & 1.55 & 0.92 & 8.35 & 0.5 & 1.6$\pm$0.3,$<$0.5 \nl
162739-244316 & GY315,YLW45,IRS51 & 24\ 37.97 & 36\ 37.0 & \nodata & G5-K7 & \nodata & \nodata & $\sim9.5$ & \nodata & 4.70 & 3.49 & 8.93 & 1-3 & 2.0$\pm$0.6 \nl
162740-242205 & GY319,SR9,EL34,ROXs29 & 24\ 38.83 & 15\ 26.4 & K4(GM),K7(CK) & K5-M2 & K8 & 3955 & 0.0 & 2.0 & 0.77 & 0.51 & 7.20 & 0.75 & 0  \nl
 & IRS52,DoAr34,WSB54 &  &  & K4/K6(BA) &  &  &  &  &  &  &  &  &  &  \nl
162742-243849 & GY326 & 24\ 40.96 & 32\ 10.9 & M4(WGM) & \nodata & M4 & 3180 & 3.2 & 0.36 & 1.78 & 0.88 & 10.59 & \nodata & \nodata  \nl
162746-242323 & GY351,VSSG13,EL35 & 24\ 45.25 & 16\ 45.2 & \nodata & M0III & b & \nodata & \nodata & \nodata & 2.20 & 1.12 & 7.29 & \nodata & 0 \nl
162749-242540 & GY372,VSSG14,EL36 & 24\ 48.37 & 19\ 02.5 & A7(GM),A7:(M) & $<$F8 & \nodata & \nodata & \nodata & \nodata & 1.20 & 0.82 & 7.32 & 0 & 0 \nl
162751-243145 & GY378,YLW52,IRS54 & 24\ 50.26 & 25\ 07.6 & \nodata & ? & \nodata & \nodata & \nodata & \nodata & 3.13 & 2.63 & 10.87 & $>$2 & 4.3$\pm$0.4 \nl
162755-242619 & GY400,SR10 & 24\ 53.97 & 19\ 41.0 & M1.5(CK) & K6-M2 & M0 & 3850 & 0.0 & 0.55 & 0.50 & 0.60 & 8.74 & 0.5 & 2.5$\pm$0.3 \nl
162757-244004 & GY410 & 24\ 56.08 & 33\ 26.1 & M6(GM) & K6-M0 & K7 & 4060 & 2.7 & 0.64 & 1.72 & 0.92 & 9.78 & 0 & 0 \nl
162757-243751 & GY411 & 24\ 56.19 & 31\ 13.3 & \nodata & G8.5III & b & \nodata & \nodata & \nodata & 2.20 & 1.21 & 9.52 & \nodata & 0  \nl
162803-242632 & GY453,VSSG16,EL37 & 25\ 02.13 & 19\ 55.5 & \nodata & K2III & b & \nodata & \nodata & \nodata & 1.10 & 0.77 & 6.89 & \nodata & 0 \nl
 & IRS58 &  &  &  &  &  &  &  &  &  &  &  &  &  \nl
162809-242322 & EL38,VSSG15 & 25\ 07.71 & 16\ 45.0 & \nodata & K2-3III & b & \nodata & \nodata & \nodata & 1.52 & 0.82 & 7.12 & \nodata & 0 \nl
162816-243657 & WSB60,YLW58 & 25\ 14.88 & 30\ 20.9 & K-M(GM) & K5-M2: & K8 & 3955 & 0.7 & 0.28 & 1.16 & 0.78 & 9.43 & 0.75 & 0  \nl
162823-242241 & SR20,ROXs33,DoAr38 & 25\ 21.82 & 16\ 05.7 & G0:(BA) & K5-M2 & K8 & 3955 & 2.1 & 0.93 & 1.69 & 1.13 & 8.55 & 1 & 0 \nl
 & WSB61 &  &  &  &  &  &  &  &  &  &  &  &  &  \nl
162923-241359 & IRAS64A & 26\ 22.00 & 07\ 27.1 & \nodata & ?\tablenotemark{k} & \nodata & \nodata & \nodata & \nodata & 1.79 & 0.91 & 6.45 & ? & 0 \nl
\nodata & ROXs39 & 27\ 33.8 & 27\ 51.0 & K5(BA) & K5-K7 & K6 & 4205 & 0.0 & 1.4 & 0.66 & 0.27 & 8.00 & 0 & 0 \nl
\tablebreak
\label{tbl:oph.properties}
\enddata
\tablenotetext{a}{Sources falling within the cloud core as defined
in Figure~1 are denoted with (c).}
\tablenotetext{b}{
SR=Struve \& Rudkj{\o}bing (1949),
DoAr=Dolidze \& Arakeylan (1959),
S and GSS=Grasdalen, Strom, \& Strom (1973),
VSSG=Vrba et al.\ (1975),
VSS=Vrba, Strom, \& Strom (1976),
EL=Elias (1978),
ROX=Montmerle et al.\ (1983),
WL=Wilking \& Lada (1983),
YLW=Young, Lada, \& Wilking (1986),
WSB=Wilking, Schwartz, \& Blackwell (1987),
IRAS=Table~1 of Wilking, Lada, \& Young (1989),
IRS=Table~2 of Wilking, Lada, \& Young (1989),
LFAM=Leous et al.\ (1991),
GY=Greene \& Young (1992).
}
\tablenotetext{c}{IR spectral types of Greene \& Meyer (1995), Greene \& Lada
(1997), and Wilking, Greene, \& Meyer (1998). Optical spectral types of
Struve \& Rudkj{\o}bing (1949), Grasdalen, Strom, \& Strom (1973),
Cohen \& Kuhi (1979), Bouvier \& Appenzeller (1992), Luhman, Liebert, \& Rieke
(1997), and Mart{\'\i}n et al.\ (1998).}
\tablenotetext{d}{Background and foreground stars are indicated by ``b" and
``f".}
\tablenotetext{e}{Derived from the IR spectra in this work.}
\tablenotetext{f}{Data from Williams et al.\ (1995) used in classification.}
\tablenotetext{g}{GY20 and GY240 are each binary systems which appear
as one source in the coordinates and photometry of BKLT.}
\tablenotetext{h}{Luhman, Liebert, \& Rieke (1997).}
\tablenotetext{i}{$H$-band classification by Biscaya et al.\ (1999).}
\tablenotetext{j}{Extinction was derived by assuming that $H-K$ is not
affected signicantly by excess emission, as suggested by the $r_K\sim0$ 
measured in the spectrum.}
\tablenotetext{k}{The optical spectral type is M8-M9III while the IR spectrum
exhibits CO absorption equivalent to K4III and is very similar to spectra
of FU Ori objects.}
\tablecomments{$JHK$ photometry and coordinates are taken from BKLT.
The data for ROXs39 and SR22 are from Greene et al.\ (1994) and references 
therein. Photometry for GY31 from Comeron et al.\ (1993).
Coordinates are prefixed by $16^{\rm h}$ and $-24\arcdeg$.}
\end{deluxetable}